\documentclass[
reprint,
notitlepage,
nofootinbib,
 amsmath,amssymb,
 aps,
pra,
10pt,
tightenlines,
floatfix,
twocolumn,
]{revtex4-2}

\usepackage{comment}
\usepackage{natbib}
\usepackage{graphicx}
\usepackage{dcolumn}
\usepackage{bm}
\usepackage{xcolor}
\usepackage{qcircuit}
\usepackage{braket}
\graphicspath{ {./figures/} }
\usepackage{subcaption}
\usepackage[export]{adjustbox}
\usepackage[font=small,labelfont=bf,
  justification=justified]{caption}
\usepackage{hyperref}
\usepackage{float}
\usepackage{svg}

\usepackage[thinc]{esdiff}

\DeclareMathOperator{\Tr}{Tr}

\usepackage{amsmath}

\hypersetup{
    colorlinks,
    citecolor=blue,
    filecolor=red,
    linkcolor=magenta,
    urlcolor=blue
}

\newcommand{\expect}[1]{\langle #1\rangle}

\begin{document}{}

\title{Improving Dynamical Decoupling for Trapped-Ion QCCD Quantum Computers}

\author{William M. Watkins}
\email{wwatki11@jhu.edu}
\affiliation{%
 Quantinuum, Broomfield, Colorado 80021, USA 
}%
\author{Leigh M. Norris}
\email{leigh.norris@quantinuum.com}
\author{Ross Hutson}
\author{Maxwell Urmey}
\author{Peter Siegfried}
\author{Charles H. Baldwin}
\affiliation{%
 Quantinuum, Broomfield, Colorado 80021, USA 
}%

\date{\today}
\begin{abstract}

We examine the impact of scheduling errors on dynamical decoupling (DD) in trapped-ion quantum charge-coupled devices (QCCDs) and develop better strategies for reducing memory errors. In the QCCD architecture, qubit transport and control introduce stochastic pulse delays that impact the efficiency of DD. Using the filter-function formalism, we analyze the performance of DD in the presence of scheduling-induced timing errors, showing that they increase the sensitivity of standard DD to low-frequency fluctuations. For typical memory errors in trapped-ion platforms, we numerically demonstrate that increasing the DD pulse frequency beyond 2 Hz is generally counterproductive due to scheduling-induced timing errors. Furthermore, we introduce a real-time DD protocol that inserts refocusing pulses opportunistically during idle periods. We demonstrate our methods on Quantinuum H2-1 with Ramsey delay-type experiments.
\end{abstract}

\maketitle

\section{Introduction}
In trapped-ion QCCD quantum computers, a key factor limiting performance are memory errors arising from low-frequency magnetic field fluctuations~\cite{Moses_2023, ransford2025helios98qubittrappedionquantum}. Dynamical decoupling (DD) is a standard technique to extend the coherence of qubits in the presence of such errors by applying single-qubit pulses at fixed time intervals~\cite{Viola_1998, Viola_1999, viola_1999_universal_control, viola_2003}. 
However, DD performance can be degraded by \emph{timing errors}, deviations between the ideal pulse placement and the pulse time realized in practice \cite{biercuk_optimized_2009, paz-silva_dynamical_2016, annealing_DD}. 
In QCCD systems, qubits (encoded in trapped-ions) are physically transported between storage zones, where qubits wait and acquire memory errors, and gate zones, where quantum operations like single-qubit pulses are applied. This limits the timing precision of single-qubit pulses since transport between different areas takes a non-negligible amount of time and compilation must schedule these pulses between operations on many qubits. 
This compilation- and transport-induced timing error, which we term \emph{scheduling error}, greatly impacts the efficacy of DD in suppressing memory error.

In this work, we analyze the impact of scheduling error on DD and develop robust DD strategies for QCCD architectures. The filter function (FF) formalism provides a unifying framework for understanding DD as a frequency-domain noise-filtering problem, enabling the systematic design and optimization of pulse sequences matched to experimentally relevant noise sources~\cite{Biercuk_2011, Green_2013, Hangleiter_2021, Cerfontaine_2021}. From the FF formalism, we derive expressions for the system dynamics averaged over stochastic timing perturbations and develop a quantitative framework to analyze the effects of scheduling errors on DD. 
This analysis reveals a counterintuitive result; for low-frequency noise sources like memory errors, a DD sequence containing \emph{fewer} pulses is preferable to a higher-order sequence with greater pulse number. To combat low-frequency noise in the presence of scheduling error, we introduce a real-time DD protocol that opportunistically inserts DD pulses during idle windows determined at execution time, rather than at compile time. This approach, which is particularly well-suited to QCCD architectures, tracks the accumulation of scheduling error and selects the timing of subsequent pulses to cancel it. 

This paper is organized as follows.
Section~\ref{sec:ctrl_noise_model}  introduces control and noise models relevant to trapped-ion QCCDs with memory error represented as a  classical, stochastic process. This section additionally provides a review of DD and cumulant expansion methods to model the dynamics of a trapped-ion qubit undergoing dephasing from memory error. In Sec.~\ref{sec::timing_errors}, we use the FF formalism to develop an analytic model for the dynamics of a trapped-ion qubit subject to both dephasing from memory error and stochastic variations in DD pulse placement due to scheduling error. This section presents our key results,  including analyses that use our analytic model as well as numerical simulations to demonstrate the impact of scheduling errors on DD performance. In Sec.~\ref{sec:real_time_dd}, we present a protocol to opportunistically insert dynamical decoupling pulses in real-time during circuit execution. To improve the performance of DD, Sec. ~\ref{sec:accounting} introduces an approach that tracks the displacements of earlier pulses and adjusts the timing of subsequent pulses to partially cancel the accumulation of scheduling error. Section~\ref{sec::experimental_results} experimentally demonstrates the effectiveness of our DD strategies in the presence of scheduling error on a trapped-ion QCCD.  We conclude in Sec.~\ref{sec:conclusion} with a summary of our main findings and an outlook on extending these ideas to broader classes of noise models.

\section{Trapped-Ion System\label{sec:ctrl_noise_model}}
In trapped-ion QCCD quantum computers, ions encode qubits and are transported between storage regions and gate zones with programmable dc electrodes. Scheduling errors result from the time required to transport an ion from a storage region to a gate zone as well as the need to coordinate transport and gating on multiple ions. For example, in this work, we demonstrate several methods on Quantinuum H2-1, a 56-qubit quantum computer arranged in a ring with four gate zones~\cite{Moses_2023}. In H2-1, the qubit is encoded in the ground hyperfine clock states of $^{171}$Yb$^+$ with $\ket{0}\equiv\ket{2S_{1/2}\, F=0, M_z=0}$ and
$\ket{1}\equiv\ket{2S_{1/2}\, F=1, M_z=0}$. The quantization axis is set by a large bias magnetic field $B$ along $z$, which also lifts the degeneracy of the $F=1$ hyperfine ground states. Single-qubit gates, such as the $\pi$-pulses we employ for DD, are realized in the gate zones via stimulated Raman transitions.

Memory errors are caused by fluctuations in the bias magnetic field and occur when qubits are idle, either during ion transport or delays while other ions gate. There are two possible types of magnetic field fluctuations: (1) transverse to $z$, which induce transitions between states in the same ground hyperfine manifold, generating leakage outside the qubit subspace, and (2) longitudinal along $z$, which lead to qubit dephasing. Previous testing and analysis strongly suggests that the second type (fluctuations along the bias field direction $z$) dominate memory errors at long times~\cite{Moses_2023, ransford2025helios98qubittrappedionquantum}, so we focus on that contribution throughout. We denote longitudinal fluctuations of the bias field by $b_z(t)$.  Because the hyperfine clock states have vanishing spin projection along $z$, they are insensitive to the linear (first-order) Zeeman shift induced by $b_z(t)$. At second order, however, the hyperfine clock states pick up a time varying frequency shift $\propto Bb_z(t)$ due to the quadratic Zeeman effect. The low-frequency dephasing induced by $b_z(t)$ and the quadratic Zeeman effect is the primary contributor to memory error.

\subsection{Open Quantum Systems Model}
In this study, we take an open quantum systems approach to model the effects of memory error and DD on a trapped-ion qubit. Because leakage due to transverse magnetic field fluctuations or single qubit gates is negligible compared to the effects of memory error over our timescale of interest, we treat the multilevel ion as an effective qubit. In a frame co-rotating with the qubit energy splitting along $\sigma_z$, the dynamics of the ion are generated by
\begin{align}
H(t)=H_{E}(t)+H_{0}(t).
\end{align}
The \emph{error Hamiltonian} $H_{E}(t)$ represents the interaction between the ion and noise sources in its environment, principally memory error in our regime of interest. The control Hamiltonian $H_{0}(t)$ generates the pulse sequences required for DD. In this section, $H_{E}(t)$ is stochastic and $H_{0}(t)$ is deterministic, as detailed below. Later, in Sec. \ref{sec::timing_errors}, the assumption of deterministic control will be relaxed as we consider the uncertainty introduced by scheduling error. 

\subsubsection{Memory Error Model}
The interaction between the ion and low-frequency magnetic field fluctuations along $z$ produces the error Hamiltonian
\begin{align}\label{eq::HE}
    H_E(t) = \frac{1}{2}\beta(t)\sigma^z,
\end{align}
where $\beta(t)\propto Bb_z(t)$ represents the noise that generates memory error.
In our open quantum systems model, $\beta(t)$ is a classical, temporally correlated, stochastic process. Similar open quantum systems models have been used previously to treat microwave amplitude and phase noise in trapped-ion systems~\cite{Soare_2014, frey_2020}. The error Hamiltonian $H_E(t)$ represents a single-axis dephasing interaction. In Sec~\ref{sec::experimental_results}, we show experimental results for Quantinuum H2-1 indicating that sources of noise transverse to $\sigma^z$ negligibly impact dynamics.

We assume that $\beta(t)$ is a Gaussian process, in which case the noisy dynamics of the trapped-ion qubit are determined by the mean and autocovariances of $\beta(t)$. Under the assumption that $\beta(t)$ is wide-sense stationary, the mean is constant $\langle{\beta(t)}\rangle_{\rm{mem}}\equiv m$ and the autocovariances take the form $\langle{[\beta(t_1)-m][\beta(t_2)-m]}\rangle_{\rm{mem}}\equiv \text{acov}(t_1-t_2)$. Here, $\langle\,\cdot\,\rangle_{\rm{mem}}$ denotes the ensemble average over realizations of $\beta(t)$. The noise power spectral density (PSD), which represents noise autocovariances in frequency space, is defined as $S(\omega)=\int^\infty_\infty dt\, e^{-i\omega t}\text{acov}(t)$. As shown in Fig.~\ref{fig:psd}, the noise responsible for memory error is generally concentrated at low frequency with a $S(2\pi f)\sim 1/f^2$ spectral dependence for $f\gtrsim 10^{-3}$ Hz. Though memory error is our primary focus, both the DD strategies we develop and our dynamical model apply to low frequency noise processes more broadly, including $1/f$ noise.

\subsubsection{Control Model}
In Quantinuum H2-1, stimulated Raman transitions are capable of generating single-qubit rotations about arbitrary axes in the $xy$-plane. Because the noise responsible for memory error is predominantly single-axis dephasing along $\sigma^z$, it is sufficient for us to consider rotations along $x$. The control Hamiltonian, therefore, takes the form
\begin{align}
H_0(t)={\frac{1}{2}}\Omega(t)\sigma^x, 
\end{align}
where $\Omega(t)$ is the effective Rabi rate of the Raman transition.

For the DD we consider, $H_0(t)$ generates
a sequence of $N$ $\pi$-pulses about $\sigma^x$ applied at times $t_1, t_2, \cdots, t_N$. These pulses invert the spin of the qubit along $z$, causing the low frequency components of $\beta(t)$ in Eq. \eqref{eq::HE} to interfere destructively, thereby suppressing dephasing. 
In this work, we treat the $\pi$-pulses as perfect and instantaneous. The assumption of perfect $\pi$-pulses holds to good approximation, as single qubit gate infidelities are on the order of $10^{-5}$ and the expected infidelity contribution from memory error over a typical idle time is $10^{-4}$~\cite{Moses_2023}. 
The instantaneous pulse approximation generally holds when the duration between $\pi$-pulses is much larger than the pulse duration. This condition is satisfied in our system, as typical idling times are $\sim 50\ \rm{ms}$ and pulse durations are $\sim 10\ \mu\rm{s}$~\cite{Moses_2023}. Under the assumption of perfect, instantaneous pulses, the control Hamiltonian takes the idealized form $H_0(t)=\pi\sum_{n=1}^N\delta(t-t_n)\sigma^x/2$. The corresponding control propagator is
\begin{align}
    \label{eqn:dd_unitary}
    U_0(t,0) = &\,\mathcal{T}_+e^{-i\int_0^t ds H_0(s)}\\\notag
    = &\,\begin{cases}
        I, & t\in[t_{j-1},t_j),\; j\leq N\;\textrm{odd}, \\
        \sigma^x, & t\in[t_{j-1},t_j),\; j\leq N\;\textrm{even},\\
        \sigma^x, &t\in[t_N,T],\; N\;\textrm{odd},\\
        I, &t\in[t_N,T],\; N\;\textrm{even},\\
    \end{cases}
\end{align}
where $T$ is the total duration of the DD sequence, $t\in[0,T]$, and $j\in\{1,\ldots,N\}$.
Note that the pulse times satisfy $t_0\equiv 0 < t_1 < t_2 < \cdots < t_N < T$. For now, the pulse times are ideal and deterministic. When we model the effects of scheduling error, the pulse times realized in practice will deviate from the ideal values of $t_1,\ldots,t_N$.

\subsubsection{Dynamical Decoupling Sequences\label{sec:dd}}
\begin{figure}
    \centering
    \includegraphics[width=\linewidth]{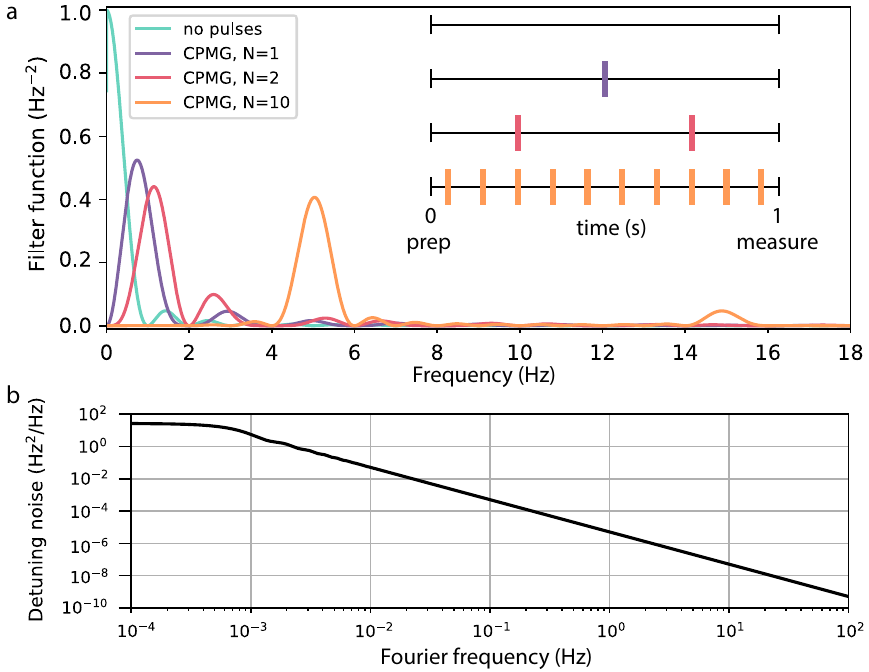}
    \caption{
        Dynamical decoupling filter functions. (a) Filter functions of CPMG pulse sequences (inset) with varying $N$ and $T=1$~s. The first harmonic of a sequence with $N$ pulses occurs at $\omega_0/2\pi= N/2T$~Hz. (b) Approximate PSD of the longitudinal magnetic field fluctuations that generate memory error. 
    }
    \label{fig:psd}
\end{figure}

Our analysis considers two paradigmatic DD sequences, Carr-Purcell-Meiboom-Gill (CPMG)~\cite{Carr_Purcell_1954, Meiboom_Gill_1958} and periodic dynamical decoupling (PDD), also known as spin-echo \cite{Hahn_1950}. A CPMG sequence of duration $T$ consists of $N$ pulses equally spaced by $T/N$ with the first pulse occurring at $T/2N$, meaning that the $j$th pulse occurs at time $t_j = T\frac{2j-1}{2N}$ for $j\in\{1,\cdots,N\}$ (Fig.~\ref{fig:psd}). In PDD, the pulses are equally spaced by $T/(N+1)$ with the first pulse occurring at $T/(N+1)$ and the
$j$th pulse occurring at time $t_j = T\frac{j}{N+1}$ for $j\in\{1,\cdots,N\}$. Observe that CPMG and PDD are equivalent for $N=1$. 

Since the goal of DD is maintaining the state of a quantum system in the presence of noise, it is desirable for a DD sequence to generate a net identity. From the control propagator in Eq. \eqref{eqn:dd_unitary}, we see that this condition is satisfied when the number of pulses $N$ is even. When $N$ is odd, we can ensure that a DD sequence generates a net identity by appending an additional $\pi$-pulse to the sequence at time $T$. Because the pulse occurs at time $T$, it has no effect on the noise suppression properties of the sequence, only on the unitary generated by the sequence.  We will assume implicitly that this final pulse has been applied in the event that $N$ is odd, ensuring $U_0(T,0)=I$. This final pulse will not be counted as one of the $N$ pulses occurring at times $t_1,\ldots,t_N$.

\subsection{Noise-Averaged Dynamics}\label{sec:error_model}
To develop an analytic model for the evolution of the qubit under memory error and DD, we evaluate the noise-averaged dynamics using a cumulant expansion \cite{kubo_1962,Cywinski2008,paz2014}. To separate the dynamical contributions of the noise and control, we first transform into the toggling frame or interaction picture associated with the control Hamiltonian $H_0(t)$. Dynamics in the toggling frame are generated by the Hamiltonian,
\begin{align}\label{eq::Rz}
\tilde{H}_E(t)=\frac{1}{2}R_z^z(t)\beta(t) \sigma^z,
\end{align}
where the control matrix element \cite{viola_1999_universal_control, Green_2013, paz_silva_2019} 
\begin{align}
\label{eqn:switching_function}
    R^z_z(t) = \frac{1}{2}\Tr\Big[ U_0(t,0) \sigma^z U_0^\dagger(t,0) \sigma^z \Big]
\end{align}
changes sign from $\pm 1$ to $\mp 1$ each time a $\pi$-pulse is applied. In this frame, the effect of dynamical decoupling is particularly clear; the sign changes induced by the $\pi$-pulses serve to average away the contribution of $\tilde{H}_E(t)$, provided $\beta(t)$ fluctuates at a frequency lower than that of the applied pulses. Evolution of the qubit in the toggling frame is related to evolution in the rotating frame by $U(T,0) = U_0(T,0)\tilde U_E(T,0)$, where $U(T,0)=\mathcal{T}_+e^{-i\int_0^T dt[H_E(t)+H_0(t)]}$ is the rotating-frame propagator and $\tilde U_E(T,0) = \mathcal{T}_+e^{-i\int^T_0 dt \tilde H_E(t)}$ is the toggling-frame propagator. Note that the two frames are equivalent when $U_0(T,0)=I$, a condition we enforce by appending a final pulse at time $T$ to a DD sequences when $N$ is odd.

\subsubsection{Control Filter Functions}\label{sec:FF}

The effectiveness of a DD sequence in suppressing noise at frequency $\omega$ is determined by the control FF, $\mathcal{F}(\omega,T)$. The control FF, which is the frequency domain representation of a time-domain control sequence, depends on the finite-time Fourier transform of the control matrix element,
\begin{align}\label{eq::FF}
\mathcal{F}(\omega,T)=\Bigg|\int_0^Tdt\,e^{-i\omega t}R_z^z(t)\Bigg|^2.
\end{align}
The decoherence rate of a qubit subject to dephasing noise is proportional to an overlap integral between the noise PSD and the control FF, $\int_0^\infty d\omega\, \mathcal{F}(\omega,T) S(\omega)$. Suppressing a noise source with PSD $S(\omega)$, thus, amounts to a filter design problem in which the pulse times of a DD sequence are selected to minimize the frequency-domain overlap between $\mathcal{F}(\omega,T)$ and $S(\omega)$ ~\cite{Biercuk_2009_opt_dd, Biercuk_2011}. A DD sequence is called \emph{ideal} if the DC $(\omega=0)$ component of the FF vanishes, i.e., $\mathcal{F}(0,T)=0$  or, equivalently, $\int_0^Tdt\,R_z^z(t)=0$. When a DD sequence is ideal, the DC component of $S(\omega)$ is perfectly suppressed. One
can verify that CPMG is an ideal sequence, while PDD is ideal only for odd $N$.

From Eqs. \eqref{eq::Rz} and \eqref{eq::FF}, it follows that the control FF generated by a DD sequence with duration $T$ and pulses at times $\{t_1,\ldots,t_N\}$ is
\begin{align}
    \label{eqn:dd_ff}
    \mathcal{F}(\omega,T) = \frac{1}{\omega^2}\Big| 1 + (-1)^{N+1} e^{i\omega T}+ 2\sum^N_{j=1} (-1)^j e^{i \omega t_j} \Big|^2 .
\end{align}
This expression can be interpreted as a sum of phasors with alternating sign. Constructive interference occurs at angular frequencies that satisfy $\omega (t_{j+1}-t_j) \approx \pi \;(\mathrm{mod}\ 2\pi$), compensating the sign alternation. 
For a CPMG sequence in which neighboring pulses are separated by a duration $T/N$, for example, the control matrix element becomes a square-wave modulation function with fundamental frequency $\omega_0 = \pi N/T$. Consequently, the FF exhibits peaks at the odd harmonics $\omega = (2k+1)\omega_0 $ for $k\in\mathbb{Z}$ and a dominant peak at the first ($k=0$) harmonic, which are visible in Fig.~\ref{fig:psd}. As $N$ increases and the spacing between neighboring pulses in a CPMG sequence becomes smaller, the spectral support of the FF shifts to higher frequency. The ``stop-band" or range of frequencies starting at $\omega=0$ where the FF is small also extends out to higher frequencies.

\subsubsection{Cumulant Expansion}
Because $\tilde{H}_E(t)$ is purely dephasing, the coupling of the qubit to the noise causes decoherence while leaving populations invariant. Ensemble-averaged over realizations of $\beta(t)$, the qubit density matrix at time $T$ takes the form
\begin{align}
\langle{\rho(T)}\rangle_{\rm{mem}} =&\, \langle \tilde U_E(T,0) \rho(0) \tilde U_E^\dagger(T,0) \rangle_{\rm{mem}}\label{eq::density_op}\\\notag
=&\, \begin{pmatrix}
        \rho_{00}(0) & \langle{e^{-i\phi(T)}}\rangle_{\rm{mem}}\,\rho_{01}(0) \\
        \langle{e^{i\phi(T)}}\rangle_{\rm{mem}}\,\rho_{10}(0) & \rho_{11}(0)
    \end{pmatrix},
\end{align}
where $\phi(T)=\int_0^TdtR_z^z(t)\beta(t)$ is the accumulated phase. Since $\phi(T)$ is a random variable, $\langle{e^{i\phi(T)}}\rangle_{\rm{mem}}$ is a characteristic function that can be expanded in terms of cumulants. The characteristic function of a random variable $X$ can be expressed as
\begin{align}
\langle e^{-iX}\rangle = \exp\left[\;\sum_{k=1}^\infty \frac{(-i)^k}{k!} \langle X^k\rangle_c \right],\label{eq::char_fun}
\end{align}
where $\langle X^k\rangle_c$ is the $k$th cumulant of $X$. The first two cumulants are simply the mean and variance,
\begin{align}
&\langle X\rangle_c=\langle X\rangle,\label{eq::Cx1}\\
&\langle X^2\rangle_c=\langle \Delta X^2\rangle
=\langle X^2\rangle-\langle X\rangle^2.\label{eq::Cx2}
\end{align}
If $X$ is normally distributed, Eq. (\ref{eq::char_fun}) truncates exactly at $k=2$. 

Using Eqs. (\ref{eq::char_fun})-(\ref{eq::Cx2}), we can write the characteristic function of $\phi(T)$ in a similar form \cite{Cywinski2008},
\begin{align}\label{eq::char_phi}
    \langle{e^{-i\phi(T)}}\rangle_{\rm{mem}} \approx \exp\left[ -iC^{(1)}(T) -\frac{1}{2}C^{(2)}(T)\right],
\end{align}
where $C^{(1)}(T)$ and $C^{(2)}(T)$ denote the first and second cumulants of $\phi(T)$, respectively.
The first cumulant, 
\begin{align}\label{eq::C1_R}
 C^{(1)}(T)=\langle\phi(T)\rangle_\text{mem}=m\int_0^Tdt\, R_z^z(t),
\end{align}
represents coherent error the qubit accumulates along $\sigma^z$. For an ideal DD sequence with $\int_0^Tdt\, R_z^z(t)=0$, the coherent error is perfectly canceled. The second cumulant, on the other hand, is related to incoherent dephasing with $C^{(2)}(T)/2$ representing the decay rate of the qubit coherence at time $T$. The second cumulant takes the explicit form
 \begin{align}
 C^{(2)}(T)=&\,\langle\Delta \phi(T)^2\rangle_\text{mem}\notag\\=&\,\int_0^Tdt_1 \int_0^Tdt_2 \,R_z^z(t_1) R_z^z(t_2)\, \text{acov}(t_1-t_2)\notag\\\label{eq::overlap}
 =&\, \int^\infty_{0}\frac{d\omega}{\pi}\mathcal{F}(\omega,T)S(\omega).
\end{align}
In the last line, we have transformed into the frequency domain to express the second cumulant in terms of the frequency-domain overlap integral discussed in Sec. \ref{sec:FF}.
Note that the cumulant expansion in Eq. (\ref{eq::char_phi}) is exact when $\beta(t)$ is a Gaussian process, making $\phi(T)$ normally distributed. For non-Gaussian $\phi(T)$, Eq. (\ref{eq::char_phi}) holds provided the noise is sufficiently weak \cite{Cywinski2008, PhysRevLett.116.150503}.

\section{Impact of Scheduling Errors} \label{sec::timing_errors}
We now analyze the impact of scheduling errors on compile-time DD applied to an idle qubit. In the previous section, we saw that the dynamical effects of memory error and DD are captured by the qubit coherence elements, $\rho_{01}(T)\propto e^{-iC^{(1)}(T)-\frac{1}{2}C^{(2)}(T)}$. Recall that the first cumulant $C^{(1)}(T)$ represents noise-induced coherent error about $\sigma^z$, while the second cumulant $C^{(2)}(T)$ is proportional to the decay rate due to incoherent dephasing. The cumulants were derived under the assumption that noise is stochastic and control is deterministic. In this section, we model scheduling error by adding stochasticity to the control. While a DD sequence may prescribe the $j$th pulse on an idle qubit to occur at time $t_j$, scheduling error will cause the pulse to be realized at a time $\tilde{t}_j$ that is generally different in practice. Because the deviation between $t_j$ and $\tilde{t}_j$ is highly dependent on details of the compiler and the circuit being evaluated, $\tilde{t}_j$ is effectively random for an arbitrary circuit instantiation. To analytically model the effects of this additional noise source, we modify the cumulant expansion of the previous section.  Using our analytic model combined with numerical simulations, we will examine the impact of scheduling errors on both the qubit dynamics and the efficacy of DD.

\begin{figure}
    \centering
    \includegraphics[width=\linewidth]{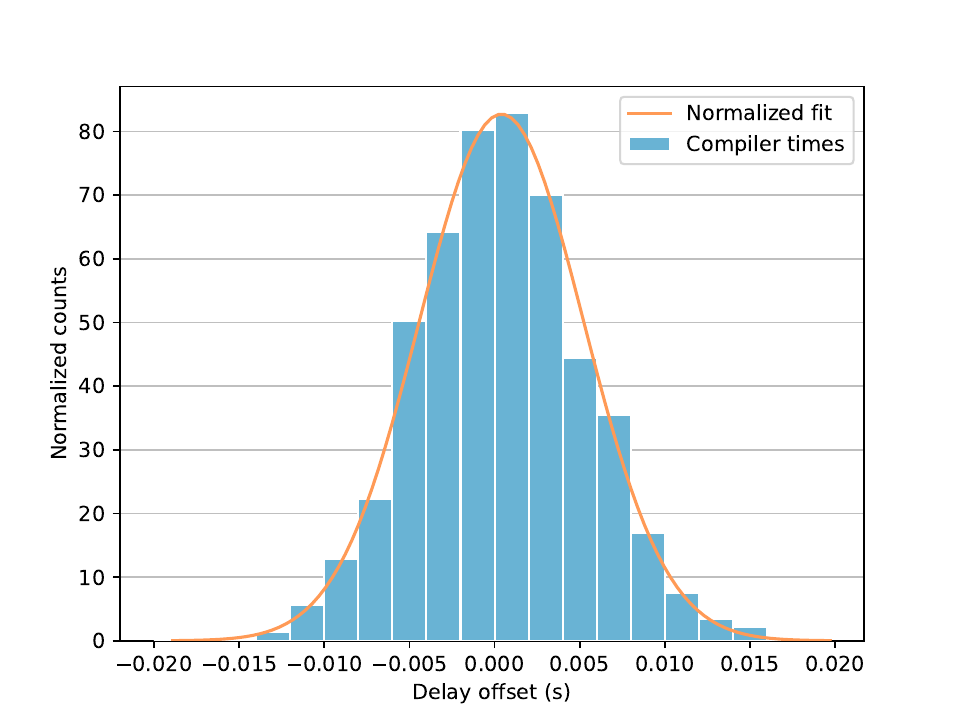}
    \caption{Empirical distribution of the timing offset $P(\epsilon)$ from Quantinuum H2 with 28 qubits. The circuits consist of single-qubit gates on each qubit iterated with two-qubit gates between each pair of qubits ($28^2/2=392$), and all repeated for four rounds. The pulses are injected at compile time and placed as close to $T/2$ as possible, where $T$ is the idle time between single-qubit gates on a given qubit. A CPMG or PDD sequence with $N=1$ prescribes that the pulse takes place at $T/2$, but because of a limited number of gate zones, the pulse is applied either earlier or later than desired. The blue bins show the offset from $T/2$ for each pulse instance during the circuit. The orange line shows a fit to a gaussian distribution with $\mu =2(1)\times10^{-4} s$, and $\sigma=4.8(1)\times10^{-3} s$. 
    }
    \label{fig:timing_errors_distr}
\end{figure}

\subsection{Scheduling error model and dynamics}
We treat scheduling errors as additive normally distributed errors on the pulse timings. If $t_j$ is the target time of pulse $j$, the actual time pulse $j$ is applied due to scheduling error is $\tilde{t}_j =t_j + \epsilon_j$, where $\epsilon_j$ is a normally distributed random variable with mean $\mu$ and variance $\sigma^2$. We assume the scheduling errors for different pulses in a DD sequence are independent and identically distributed (i.i.d.), so that $\text{cov}(\epsilon_j\epsilon_k)\equiv\langle\epsilon_j\epsilon_k\rangle_\text{sch}-\langle\epsilon_j\rangle_\text{sch}\langle\epsilon_k\rangle_\text{sch}=\sigma^2\delta_{j,k}$, where $\langle\,\cdot\,\rangle_\text{sch}$ denotes the expected value over the distribution of scheduling error. A nonzero mean ($\mu\neq 0$) indicates a bias in the scheduling errors, shifting all pulse times earlier or later on average by $\mu$. This model is supported by empirical scheduling error distributions obtained from the Quantinuum H2 compiler.
Figure \ref{fig:timing_errors_distr} shows the empirical distribution of the difference between the actual and ideal pulse time $\epsilon_1=\tilde{t}_1-t_1$ for the first pulse of a $N=1$ CPMG/PDD sequence. Observe that the distribution is well-approximated as Gaussian with $\mu=2(1)\times 10^{-4}$ s and $\sigma=4.8(1)\times 10^{-3}$ s. 

Our scheduling error model presents some subtleties relating to the ordering of the displaced $\pi$-pulses $\tilde{t}_j=t_j+\epsilon_j$. While the ideal pulse times are ordered such that $0<t_1 < \cdots < t_{N}< T$, the addition of scheduling error $\epsilon_j$ can alter the ordering or cause the perturbed pulse times to fall outside the idle time interval $[0,T]$. Note that the probability of such events is negligible in experimentally relevant regimes where $|\mu|$ and $\sigma$ are much smaller than $T/N$. For completeness, however, our analysis considers regimes where $|\mu|$ and $\sigma$ are greater than or on the same order as $T/N$. To address the possibility of altered ordering, we define $\tilde t_{(j)}$ as the $j$th-smallest value in $\{\,\tilde{t}_j=t_j+\epsilon_j\,|\,j=1,\ldots, N\}$, so that the pulses are applied in chronological order at times $\tilde t_{(1)}\leq\ldots\leq\tilde t_{(N)}$. Note that $\tilde t_{(j)}$ is similar but not equivalent to the order statistic of a random sample. The $j$th order statistic is defined as the $j$th-smallest value of an i.i.d. random sample. The $\tilde t_{j}$ are not identically distributed, however, because they have different means, $\langle\tilde t_{j}\rangle_\text{sch}=t_j+\mu$. A pulse time $\tilde t_{j}$ falling outside the idle time interval signifies that pulse $j$ cannot be scheduled. On a QCCD, this could occur, for example, because all gate zones are in use. If an even number of pulses fall outside the idle time interval, the unitary generated by the DD sequence will remain $U_0(T,0)$, though the filtering properties of the sequence could change substantially. If an odd number of pulses fall outside the idle time interval, the unitary generated by the sequence will be $\sigma_x  U_0(T,0)$. In the later case, we can apply an additional $\pi$-pulse when the qubit becomes active at time $T$ to recover the target unitary $U_0(T,0)$. This final pulse has no scheduling error, as the qubit enters the gate zone when the idle time ends at $T$.

\subsection{Scheduling Error Cumulants}\label{sec::sch_error_cumulants}
In Sec.~\ref{sec:error_model}, we used a cumulant expansion to evaluate the noise-averaged qubit dynamics. Because the pulse times $\{\tilde{t}_1,\ldots,\tilde{t}_N\}$ are now random variables, 
we must also average the qubit dynamics over the distribution of scheduling errors. The resulting qubit density operator, $\langle\langle{\rho(T)}\rangle_{\rm{mem}}\rangle_{\rm{sch}}$, takes the same form as Eq. (\ref{eq::density_op}) with the characteristic function now averaged over the scheduling errors in addition to the memory error,
\begin{align}
    \langle\langle e^{-i\phi(T)} \rangle_{\rm{mem}} \rangle_{\rm{sch}}\approx &\, \langle e^{-iC^{(1)}(T)-\frac{1}{2}C^{(2)}(T)} \rangle_{\rm{sch}}\notag\\ \approx &\,e^{K_\text{sch}^{(1)}(T)+K_\text{sch}^{(2)}(T)}.
\end{align}
Note that $C^{(1)}(T)$ and $C^{(2)}(T)$ are random variables due to their dependence the $\{\tilde{t}_1,\ldots,\tilde{t}_N\}$.
We can therefore define the ``scheduling error cumulants" $K_\text{sch}^{(1)}(T)$ and $K_\text{sch}^{(2)}(T)$ as the first and second order cumulants of $-iC^{(1)}(T)-\frac{1}{2}C^{(2)}(T)$, obtained from Eqs. (\ref{eq::char_fun})-(\ref{eq::Cx2}) with $X=-iC^{(1)}(T)-\frac{1}{2}C^{(2)}(T)$. Preserving terms up to second order in $\beta(t)$, we find
\begin{align}\label{eq::K1}
K_\text{sch}^{(1)}(T)=&\,-i\big\langle C^{(1)}(T) \big\rangle_{\rm{sch}}- \frac{1}{2} \big\langle C^{(2)}(T) \big\rangle_{\rm{sch}},\\\label{eq::K2}
K_\text{sch}^{(2)}(T)=&\,
-\frac{1}{2}\Big[ \big\langle C^{(1)}(T)^2 \big\rangle_{\rm{sch}} - \big\langle C^{(1)}(T) \big\rangle_{\rm{sch}}^2 \,\Big].
\end{align}
The first term in $K_\text{sch}^{(1)}(T)$ represents coherent error along $\sigma_z$ that is no longer fully canceled by DD on account of bias in the scheduling error. This occurs because $\int_0^Tdt\, \langle R_z^z(t)\rangle_\text{sch}\neq 0$ when $\mu\neq 0$, even if the DD sequence is ideal. The second term in $K_\text{sch}^{(1)}(T)$ is a contribution from incoherent dephasing due to stochastic fluctuations in $\beta(t)$.
The second cumulant, $K_\text{sch}^{(2)}(T)$, represents additional incoherent dephasing resulting from the variance of $C^{(1)}(T)$, which is nonzero when  $\sigma^2\neq 0$. Both the coherent component of $K_\text{sch}^{(1)}(T)$ and incoherent error in $K_\text{sch}^{(2)}(T)$ represent qualitatively new dynamical contributions arising from scheduling error.

\subsection{Coherent Memory Error\label{ssec:coherent_env}}

To start, we examine the impact of scheduling errors under the assumption that the memory error is purely coherent, i.e., $\beta(t)$ is deterministic and constant so that $\beta(t)=\langle\beta(t)\rangle_\text{mem}=m$. 
We specialize to CPMG, though we expect our findings to apply qualitatively to other DD sequences.
In the coherent regime, $C^{(2)}(T)=0$.
Consequently, the scheduling error cumulants depend only on $C^{(1)}(T)$,
\begin{align}
&K_\text{sch}^{(1)}(T)=-i\langle C^{(1)}(T)\rangle_\text{sch},\label{eq::K1_coherent}\\
&K_\text{sch}^{(2)}(T)=-\frac{1}{2}\rm{var}[C^{(1)}(T)].\label{eq::K2_coherent}
\end{align}
Recall that $\langle C^{(1)}(T)\rangle_\text{sch}$
represents coherent error along $\sigma^z$ that arises when $\mu\neq 0$, 
while $\rm{var}[C^{(1)}(T)]$ represents incoherent dephasing that occurs due to the scheduling error variance $\sigma^2$. Counterintuitively, even in the presence of a purely coherent environment, scheduling errors generate incoherent dephasing.

\begin{figure}
    \includegraphics[width=.48\textwidth,left]{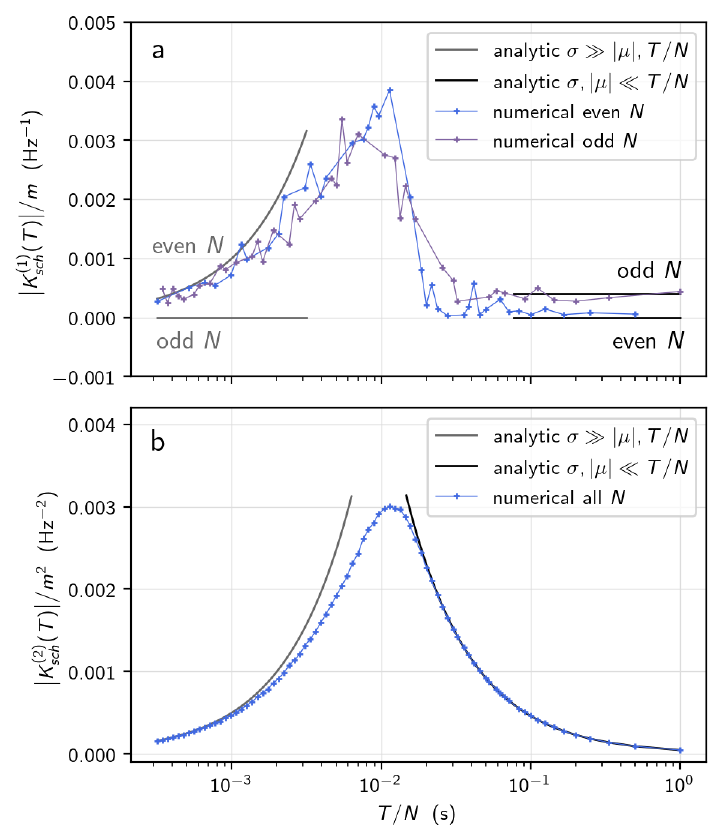}
    \caption{Scheduling error cumulants for CPMG plotted over a range of average pulse spacings $T/N$. Here, the CPMG pulse number $N$ is varied with fixed $T=1$ s, $\mu=2\times 10^{-4}$ s, and $\sigma=4.8\times 10^{-3}$ s. 
    The magnitudes of the first scheduling error cumulant $|K_\text{sch}^{(1)}(T)|$ (a) and the second scheduling error cumulant $|K_\text{sch}^{(2)}(T)|$ (b) are computed numerically for each $T/N$. In the limits where the pulse spacing $T/N$ is either much smaller or much larger than $\sigma$, the numerically computed cumulants converge to the analytic expressions given in Sec. \ref{ssec:coherent_env}. When $\sigma\gg |\mu|,\,T/N$, the numerical value of $|K_\text{sch}^{(1)}(T)|$ lies between the analytic expression for odd and even $N$ due to pulses falling outside the idle time interval $[0,T]$.
    }
    \label{fig:error_statistics}
\end{figure}

A key figure for determining the dynamical effects of the coherent and incoherent error contributions is the magnitude of $\mu$ or $\sigma$ relative to $T/N$, the expected duration between pulses in a CPMG sequence. We first consider the regime where the magnitude of the mean and standard deviation are small compared to the expected duration between pulses, i.e., $|\mu|,\,\sigma\ll T/N$.
In this regime, we take $\tilde t_{(j)}=\tilde t_j$ since it is unlikely that pulse ordering is altered or that displaced pulses fall outside the time boundary. From Eq. \eqref{eq::C1_R}, the coherent error contribution is
\begin{align}
    \label{eqn:cumulant_tj_epsilon}
    C^{(1)}(T) &= m \bigg[ 2\sum^{N}_{j=1}(-1)^{j-1} \big( t_j+\epsilon_j \big) + (-1)^N T \,\bigg] \\
    &= 2 m \sum^{N}_{j=1}(-1)^{j-1} \epsilon_j.\nonumber
\end{align}
Here, we have used $\int_0^T dt R_z^z(t)=2\sum^{N}_{j=1}(-1)^{j-1}t_j+(-1)^N T=0$, which holds
for an ideal DD sequence. It follows that
\begin{align}\label{eqn:coherent_error_stats_mean}
K_\text{sch}^{(1)}(T)=&\,-i\,2 m \sum^{N}_{j=1}(-1)^{j-1}\langle \epsilon_j\rangle_\text{sch}\\\notag=&\,
\begin{cases}
-2i\,m\mu, & N\;\text{odd},\\
0, & N\;\text{even},
\end{cases}
\\\notag\\
K_\text{sch}^{(2)}(T)=&\, \label{eqn:coherent_error_stats_var}
    -\frac{1}{2}\sum^{N}_{j=1}\textrm{var}[\,2 m (-1)^{j-1} \epsilon_j]  \\\notag=&\, -2 m^2 \sigma^2 N.
\end{align} 
The coherent error is canceled when the number of pulses is even, while the dephasing rate $2 m^2 \sigma^2 N$ grows linearly with the pulse number. Each pulse effectively injects incoherent dephasing due to scheduling error. Consequently, a large pulse number does not minimize the error. The optimal pulse number will instead be $N=1$ or $N=2$, depending on the relative magnitudes of $\mu$, $\sigma$, and $m$.

Next, we examine the case where the pulse number is large, $N\gg 1$, and the standard deviation of the scheduling error is greater than the magnitude of the mean and the average pulse spacing, $\sigma \gg |\mu|,\, T/N$. In the large $N$ limit, the $\tilde{t}_j$ are so tightly packed relative to their spread that they are approximately uniformly distributed over the interval $[0,T]$. In Appendix \ref{sec::pulse_time_distributions}, we show numerically computed pulse time distributions in both the $\sigma \gg |\mu|,\, T/N$ for $N\gg 1$ and $|\mu|,\,\sigma\ll T/N$ limits. Because the scheduling error is likely to change the order of the pulses when $\sigma \gg T/N$, $\tilde t_{(j)}\neq \tilde t_j$ in general. Because the $\tilde{t}_j$ are all approximately uniformly distributed, they are also  i.i.d. in this regime, meaning that the  $\tilde{t}_{(j)}$ are true order statistics. 

We can determine the scheduling error cumulants by leveraging the properties of order statistics.
For a set of $N$ i.i.d. samples from the uniform distribution on $(0,1)$, the $j$th order statistic is a random variable with beta distribution, $x_{(j)}\sim \mathrm{Beta}(j, N+1-j)$. We assume that $\sigma$ is much greater than $T/N$ but  sufficiently small so that the number of $\tilde{t}_j$ falling outside the idle time interval is much less than $N$. In this case,
we can approximate the number of $\tilde{t}_j$ contained in $(0,T)$ as $N$, implying that $\tilde t_{(j)} \approx T x_{(j)}$. The moments of the $\tilde t_{(j)}$ follow from the moments of the $x_{(j)}$, namely,
\begin{align}
    \label{eqn:ordered_timings_mean}
    &\langle \tilde{t}_{(j)}\rangle_\text{sch} = 
        \frac{jT}{N+1},\\
    &\text{cov}[\tilde{t}_{(j)}\tilde{t}_{(k)}]= T^2 \frac{j(N+1-k)}{(N+1)^2(N+2)},\; k\geq j.
\end{align}
Equation \eqref{eqn:ordered_timings_mean} shows that the expected DD sequence under scheduling error is equivalent to PDD with $N$ pulses. The scheduling error cumulants follow from the moments of the order statistics,
\begin{align}\label{eq::K1_order_stat}
K_\text{sch}^{(1)}(T)=&\,-i\,m\bigg[(-1)^NT-2\sum_{j=1}^N(-1)^j\langle\tilde{t}_{(j)}\rangle_\text{sch}\bigg]\\
=&\,-i\,m\, T\begin{cases}
0,&N\;\text{odd},\\
\frac{1}{N+1},&N\;\text{even},\notag
\end{cases}
\end{align}
\begin{align}\label{eq::K2_order_stat}
K_\text{sch}^{(2)}(T)=&\,-2m^2\sum_{j,k=1}^N(-1)^{j+k}\text{cov}[\tilde{t}_{(j)}\tilde{t}_{(k)}]\\\notag
=&\,-2m^2T^2\bigg[\frac{1-(-1)^N+2N(N+2)}{8(N+1)^2(N+2)}\bigg].
\end{align}
Both the coherent error captured by $K_\text{sch}^{(1)}(T)$ and the incoherent dephasing captured by $K_\text{sch}^{(2)}(T)$ tend to zero with increasing pulse number. This shows that in the limit $\sigma \gg |\mu|, T/N$, $N\gg 1$, the pulses are so closely spaced that succeeding pulses effectively decouple the scheduling error introduced by the pulses that came before. This is a stark contrast to the case of $|\mu|,\,\sigma\ll T/N$, where the dephasing rate scaled linearly with the pulse number. 

Figure~\ref{fig:error_statistics} shows the magnitudes of the scheduling error cumulants for CPMG with varying pulse number $N$ and fixed $T$. The scheduling error cumulants are determined by computing $C^{(1)}(T)$ numerically via Eq. \eqref{eq::C1_R} and averaging over 40,000 scheduling error realizations for each CPMG sequence. The average pulse spacing of the CPMG sequences $T/N$ varies from approximately $.07\sigma$ to $ 108\sigma$. When $\sigma\gg T/N$, $|K_\text{sch}^{(1)}(T)|$ and $|K_\text{sch}^{(2)}(T)|$ converge to to the analytic expressions in Eqs. \eqref{eq::K1_order_stat} and \eqref{eq::K2_order_stat}, both tending to zero as $T/N$ decreases relative to $\sigma$.  Note that $|K_\text{sch}^{(1)}(T)|$ falls between the analytic expressions for odd and even $N$ due to pulses falling outside the idle time interval when $\sigma\gg T/N$, meaning that the actual number of pulses in the sequence differs from $N$. In the opposite $\sigma\ll T/N$ regime, $|K_\text{sch}^{(1)}(T)|$ and $|K_\text{sch}^{(2)}(T)|$ converge to to the analytic expressions in Eqs. \eqref{eqn:coherent_error_stats_mean} and \eqref{eqn:coherent_error_stats_var}. With increasing $T/N$, $|K_\text{sch}^{(2)}(T)|$ tends to zero as there are fewer pulses to introduce incoherent dephasing. The limiting behavior of the coherent error contribution $|K_\text{sch}^{(1)}(T)|$ when $\sigma\ll T/N$, on the other hand, depends on whether the pulse number is even or odd with $|K_\text{sch}^{(1)}(T)|$ only tending to zero for even pulse numbers. In the intermediate regime around $T/N=10^{-2}\,\text{s}\approx 2\sigma$, the scheduling error cumulants go through a ``phase transition" at which $|K_\text{sch}^{(1)}(T)|$ and $|K_\text{sch}^{(2)}(T)|$ are maximal due to accumulated scheduling error from the pulses. The pulses are not yet sufficiently close together to decouple the scheduling error, as in the $\sigma\gg T/N$ limit.

\subsection{Stochastic, Temporally-Correlated Memory Error \label{ssec:stoch_env}}
Next, we consider the more general case 
of a temporally-correlated, stochastic $\beta(t)$ with a mean $m$ and PSD $S(\omega)$. In this section, we work in the physically relevant $|\mu|,\,\sigma \ll T/N$ regime, where the pulse timings are not reordered by the scheduling errors and all pulses fall within the idle time interval. Similar to the case of coherent $\beta(t)$ in the previous section, we analyze the impact of scheduling error through the scheduling error cumulants. The first term of $K_\text{sch}^{(1)}(T)$ in Eq. \eqref{eq::K1} and the expression for $K_\text{sch}^{(2)}(T)$ in \eqref{eq::K2}, which depend only on $m$, were examined in the previous section. Here, we focus on the second term of $K_\text{sch}^{(1)}(T)$ in Eq. \eqref{eq::K1}, which we can write in the frequency domain as
\begin{align}\label{eq::C2_FF_av}
    \langle C^{(2)}(T) \big\rangle_{\rm{sch}} = \int^\infty_0 \frac{d\omega}{\pi}\langle \mathcal{F} (\omega,T) \big\rangle_{\rm{sch}} S(\omega).
\end{align}
This expression is almost identical to the overlap integral in Eq. \eqref{eq::overlap}, except the FF is averaged over the distribution of timing error. Scheduling error, therefore, alters the filtering properties of a DD sequence and affects the extent to which stochastic memory error can be suppressed.

To understand the influence of scheduling error on the FF, we obtain an analytic expression for $\langle \mathcal{F} (\omega,T) \big\rangle_{\rm{sch}}$ by first expanding Eq.~\eqref{eqn:dd_ff},
\begin{align}\label{eq::FF_av1}
    &\langle \mathcal{F}(\omega,T) \rangle_{\text{sch}} = \frac{1}{\omega^2}\bigg\{\; \big\vert 1+(-1)^{N+1}e^{i\omega T}\big\vert^2 \\
    &\;\;\;\;\;\;\;+ 4\text{Re}\bigg[ \big(1+(-1)^{N+1}e^{i\omega T}\big)\sum^N_{j=1}(-1)^j \langle e^{-i\omega \tilde t_j} \rangle_{\text{sch}} \bigg] \nonumber\\
    &\;\;\;\;\;\;\;+ 4\sum^N_{j=1}\sum^N_{k=1}(-1)^{j+k} \langle e^{i\omega (\tilde t_j - \tilde t_k)} \rangle_{\text{sch}} \;\bigg\}.\notag
\end{align}
The expected values of the phasors in this expression are characteristic functions of $\tilde{t}_j$ and $\tilde{t}_j-\tilde{t}_k$ that we can evaluate using the cumulant expansion in Eq. \eqref{eq::char_fun} to second order,
\begin{align}
&\big\langle e^{\!-i\omega \tilde t_j} \big\rangle_{\text{sch}} = \exp\!\bigg\{-i\omega \langle\tilde t_j\rangle_{\text{sch}} - \frac{1}{2}\omega^2 \text{var}[\tilde t_j]\; \bigg\},\\\notag
&\;\;\;\;\;\;\;\;\;\;\;\;\;\;\;\;\;=e^{-i\omega(t_j+\mu)-\frac{1}{2}\omega^2\sigma^2}\\\nonumber\\
&\big\langle e^{i\omega(\tilde t_j-\tilde t_k)} \big\rangle_{\text{sch}} = \exp\bigg\{i\omega \langle\tilde t_j-\tilde t_k\rangle_\text{sch} \\\nonumber
&\;\;\;\;\;\;\;\;\;\;\;\;\;\;\;\;\;\;\;\;\;\;\;\;\;- \frac{1}{2}\omega^2\Big( \text{var}[\tilde t_j] + \text{var}[\tilde t_k] - 2 \text{cov}[\tilde t_j,\tilde t_k]\Big) \!\bigg\}\\\notag
&\;\;\;\;\;\;\;\;\;\;\;\;\;\;\;\;\;\;\;\;\;\;
=e^{i\omega(t_j-t_k)-\omega^2\sigma^2}
\end{align}
where $j\neq k$. Since the $\tilde t_j$ are normally distributed, these expressions are exact. Substituting the time-averaged phasors into Eq. \eqref{eq::FF_av1}, we find
\begin{align}
\label{eqn:ff_analytic_scheduling_errors}
    &\langle \mathcal{F}(\omega,T) \rangle_{\text{sch}} = \frac{1}{\omega^2}\bigg\{
    \Big\vert 1+(-1)^{N+1}e^{i\omega T}\Big\vert^2 \\
    &+ 4\text{Re}\Big[(1+(-1)^{N+1}e^{i\omega T})e^{-i\omega\mu-\frac{1}{2}\omega^2\sigma^2}\sum^N_{j=1}(-1)^j e^{-i\omega t_j} \Big] \nonumber\\\notag
    &+ 4e^{-\omega^2\sigma^2}\sum^N_{j=1}\sum^N_{k=1}(-1)^{j+k} e^{i\omega(t_j-t_k)} 
    +4N(1-e^{-\omega^2\sigma^2}) \bigg\}.
\end{align}
This FF captures the influence of scheduling error on the filtering properties of a DD sequence. 

Scheduling errors negatively impact the capability of a DD sequence to filter low frequency noise.   
Taking the $\omega\rightarrow 0$ limit of Eq. \eqref{eqn:ff_analytic_scheduling_errors} gives us the DC component of the FF,
\begin{align}\notag
    \langle \mathcal{F}(0,T) \rangle_{\text{sch}} 
    =&\, \mathcal{F}_\text{ideal}(0,T)
    \\\notag&\,-2\mu\big[(-1)^N-1\big]\int_0^Tdt\, R_{z,\,\text{ideal}}^z(t)\\\label{eq::FF_av_DC}&+ \big[(-1)^N-1\big]^2\mu^2+4N\sigma^2,
\end{align}
which is equivalent to DC component of  the ideal FF without scheduling error plus an additive bias.
Even for an ideal DD sequence with $\mathcal{F}_\text{ideal}(0,T)=0$ and $\int_0^Tdt\, R_{z,\,\text{ideal}}^z(t)=0$, the DC component of $\langle \mathcal{F}(0,T) \rangle_{\text{sch}}$ is generally non-vanishing and, furthermore, grows with the number of pulses. This is particularly problematic for noise sources with PSDs concentrated at low frequency, 
such as the PSD depicted in Fig.~\ref{fig:psd}. Equation \eqref{eq::FF_av_DC} shows that applying more pulses has the perverse effect of increasing the susceptibility of the trapped-ion qubit to low frequency noise. 
Similar to the case of coherent memory error in the physically relevant $|\mu|,\sigma \ll T/N$ regime, 
DD is most effective when fewer pulses are used. 

\begin{figure}
    \centering
    \includegraphics[width=\linewidth]{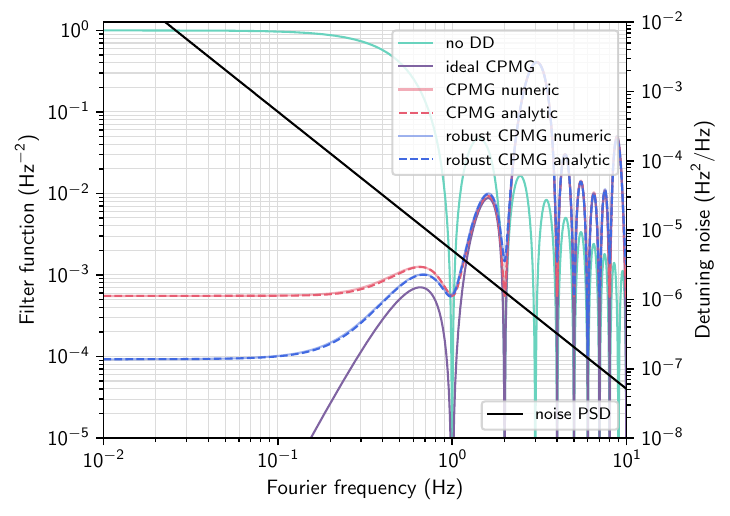}
    \caption{Effect of scheduling errors on FFs. For $T=1$ s, $\mu=2\times 10^{-4}$ s, and $\sigma=4.8\times 10^{-3}$ s, FFs are shown for free evolution without DD (cyan), CPMG with $N=6$, and robust CPMG with $N=6$. The FFs of CPMG averaged over scheduling error (pink solid/dashed) have a significantly larger DC component than the ideal CPMG FF without scheduling error (purple), which falls off quadratically in the DC regime. Scheduling-error-robust DD (blue solid/dashed) reduces the DC component of the FF by almost an order of magnitude. Numerical FFs for CPMG (pink solid) and robust CPMG (blue solid), which were generated by averaging over 1000 realizations of scheduling error, show excellent agreement with the analytic formulas in Eq. \eqref{eq::FF_av1} (pink dashed) and Eq. \eqref{eq::robust_FF} (blue dashed).
    }
    \label{fig:timing_error_ff}
\end{figure}

\begin{figure}
    \centering
    \includegraphics[width=0.435\textwidth]{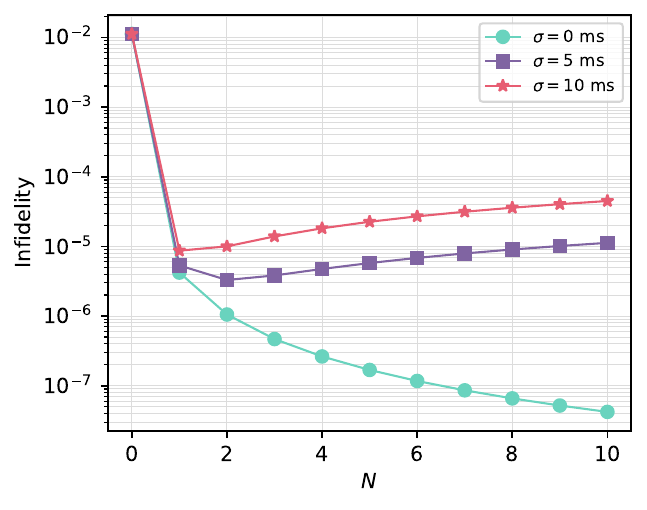}
    \caption{Infidelity versus pulse number $N$ for compile-time CPMG and free evolution. Memory error is modeled as purely stochastic ($m=0$) with the $1/\omega^2$ PSD shown in Fig. \ref{fig:psd}. For three different values of $\sigma$, $\mu=0$ s, and a fixed idle time $T=1$ s, the infidelity of the trapped-ion qubit is plotted for CPMG with $1\leq N\leq 10$ and free evolution with $N=0$ (no pulses). When $\sigma=0$ ms (cyan), the infidelity decreases with increasing $N$ since no scheduling error accompanies the pulses. When $\sigma=5$ ms (purple) and $\sigma=10$ ms (pink), infidelity is minimized at $N=2$ and $N=1$ or, equivalently, at a DD pulse frequency of 2 Hz and 1 Hz, respectively. The optimal $N$ values suppress memory error while minimizing the scheduling error that accumulates with higher pulse numbers. }
    
    \label{fig:sq_fidelity_vs_N}
\end{figure}

The impact of scheduling error on the filtering properties of CPMG with $N=6$ is illustrated in Fig.~\ref{fig:timing_error_ff}. In the presence of scheduling error, the average FF $\langle \mathcal{F}(\omega,T) \rangle_{\text{sch}}$ levels to a finite ``shelf" at low frequency, as predicted by Eq. \eqref{eq::FF_av_DC}. The finite shelf indicates that scheduling error reduces the noise suppression of CPMG at low frequencies. The FF of CPMG without scheduling error, in contrast, decays quadratically as $\omega\rightarrow 0$, which is typical of ideal DD sequences. Figure  \ref{fig:sq_fidelity_vs_N} shows the dependence of the qubit infidelity on the number of CPMG pulses $N$ for purely stochastic memory error. In absence of scheduling error, the infidelity decreases with increasing $N$ since the stop-band of the FF grows, suppressing memory error over a larger range of frequencies. In the presence of scheduling error, the increased suppression of memory error for larger $N$ is offset by the accumulation of scheduling error. The minimum infidelity instead occurs at smaller pulse numbers ($N=1$ or $N=2$), which balance the tradeoff between memory error suppression and scheduling error. For the realistic scheduling error strength of $\sigma = 5$ ms, infidelity is minimized at $N=2$ or, equivalently, a DD pulse frequency of 2 Hz.

\section{Real-time Dynamic Decoupling\label{sec:real_time_dd}}

Real-time dynamical decoupling (RTDD) is an algorithm to insert DD pulses opportunistically at runtime to suppress memory error.
Instead of specifying a DD sequence $t_1,\ldots,\,t_N$ to be injected at compile time for a known idle time $T$, in RTDD the idle time is unknown a priori, so we specify a \emph{DD threshold} $\Delta t$.
The algorithm attempts to opportunistically place DD pulses at execution time after a qubit has been idle for at least $\Delta t$.
RTDD is designed for QCCDs where circuits are compiled in real time, such as Quantinuum Helios~\cite{ransford2025helios98qubittrappedionquantum}. 

To account for the uncertainty in the idle time $T$ when circuits are compiled in real time, we can view $T$ as a random variable drawn from an \emph{idling time distribution}, $p(T)$. For a random $T$ drawn from $p(T)$ and a threshold $\Delta t\gg \sigma,\, |\mu|$, the number of pulses that can be applied to the qubit is $N= \lfloor T/ \Delta t \rfloor$. After the idle time begins, the RTDD algorithm will attempt to apply $N$ pulses spaced by duration $\Delta t$, though there will be some stochasticity in the placement of the pulses due to scheduling error. In absence of scheduling error, in the case where $T/\Delta t=N+1$, RTDD is equivalent to PDD with $N$ pulses.  

To avoid specifying the details of $p(T)$, we focus on the ``remainder time" $r=T-N\Delta t$. The remainder time, defined as the duration between the ideal placement of the $N$th pulse and $T$, is another random variable depending on $p(T)$. We take $r$ to be independent of scheduling error, so that $\text{cov}[r\epsilon_j]=0$ for $j\in\{1,\ldots,N\}$. If the threshold time is sufficiently small, $p(T)$ is flat on timescales of order $\Delta t$ and the distribution of $r$ is approximately uniform, 
\begin{align}
    r \sim \mathrm{Uniform}[0,\Delta t].
\end{align}
 Since no pulses are applied during the remainder time, the qubit accumulates memory error by an amount that depends on the remainder time duration. Real-time compilation, therefore, introduces a new element of stochasticity impacting the effectiveness of DD.

To assess the dynamical impact of real-time compilation, we follow a procedure similar to Sec. \ref{sec::sch_error_cumulants}  and use Eqs. (\ref{eq::char_fun})-(\ref{eq::Cx2}) with $X=-iC^{(1)}(T)-\frac{1}{2}C^{(2)}(T)$ to define cumulants depending on averages over both scheduling error and the remainder time. Accounting for the uncertain remainder time, the decoherence of the qubit is captured by  $\langle\langle e^{-i\phi(T)} \rangle_{\rm{mem}} \rangle_{\rm{rt}}\approx e^{K_\text{rt}^{(1)}(T)+K_\text{rt}^{(2)}(T)}$, where $\langle\,\cdot\,\rangle_\text{rt}$ denotes the expected value over the distributions of both the scheduling error and the remainder time. The ``real-time" cumulants are
\begin{align}
    \label{eqn:real_time_cumulant}
   K_\text{rt}^{(1)}(T) =&\, -i\big\langle C^{(1)}(T) \big\rangle_{\rm{rt}} - \frac{1}{2} \big\langle C^{(2)}(T)\big\rangle_\text{rt},\nonumber\\
   K_\text{rt}^{(2)}(T) =&\,
   -\frac{1}{2}\Big[ \big\langle C^{(1)}(T)^2 \big\rangle_{\rm{rt}} - \big\langle C^{(1)} (T)\big\rangle_{\rm{rt}}^2 \big], \nonumber
\end{align}
where we have preserved terms up to second order in $\beta(t)$. Similar to the scheduling error cumulants in Eqs. \eqref{eq::K1} and \eqref{eq::K2}, the first term in $K_\text{rt}^{(1)}(T)$ represents coherent error that arises from imperfect DD pulse timings due to scheduling error or the extra accumulation of memory error during the remainder time. The second term in $K_\text{rt}^{(1)}(T)$ results from incoherent dephasing due to stochastic fluctuations in $\beta(t)$. The second cumulant, $K_\text{rt}^{(2)}(T)$ sets the decay rate from additional incoherent dephasing due to fluctuations in the DD pulse times and the remainder time.

We focus first on contributions that depend on $C^{(1)}(T)$. For a RTDD sequence with pulses occurring at times $\tilde{t}_j=j\Delta t+\epsilon_j$ $(j=1,\ldots,N)$, we find
\begin{align}
    C^{(1)}(T) = &\,2m\sum^N_{j=1} (-1)^{j-1}j\Delta t + m(-1)^N N\Delta t \\\notag
    &+ 2m\sum^N_{j=1}(-1)^{j-1}\epsilon_j + m(-1)^N r.
\end{align}
The first line represents the coherent memory error accumulated during the``ideal" RTDD sequence consisting of $N$ pulses at times $t_j=j\Delta t$. The sequence perfectly cancels coherent error when $N$ is even. The second line captures additional coherent error accumulated due to imperfect pulse placement and the remainder time. By taking the mean and variance of $C^{(1)}(T)$ with respect to the scheduling error and remainder time distributions, we find
\begin{align}
   \langle C^{(1)}(T)\rangle_\text{rt}=\begin{cases} \frac{m\Delta t}{2},& N\; \text{even},\\
   \frac{m\Delta t}{2}+2m\mu,& N\; \text{odd},\end{cases}
\end{align}
and
\begin{align}
    \label{eqn:coherent_error_stats_var_real_time}
    K_\text{rt}^{(2)}(T) = -m^2\bigg( \frac{\Delta t^2}{24} + 2 \sigma^2 N \bigg).
\end{align}
Due to the remainder time, the coherent error contribution in $K_\text{rt}^{(1)}(T)$ grows linearly with $\Delta t$, while the incoherent dephasing rate represented by $|K_\text{rt}^{(2)}(T)|$ scales as $\Delta t^2$. On account of scheduling error, $|K_\text{rt}^{(2)}(T)|$ also scales linearly with $N\sim\frac{1}{\Delta t}$. This suggests the performance of RTDD can be enhanced by selecting a $\Delta t$ that optimally balances the contributions from scheduling error and the remainder time.

To obtain the second term in $K_\text{rt}^{(1)}(T)$, which is proportional to 
\begin{align}
\langle C^{(2)}(T) \big\rangle_{\rm{rt}} = \int^\infty_0 \frac{d\omega}{\pi}\langle \mathcal{F} (\omega,T) \big\rangle_{\rm{rt}} S(\omega), 
\end{align}
we take the average of the FF similar to Sec. \ref{ssec:stoch_env}. Starting from Eq.~\eqref{eqn:dd_ff}, the FF generated by RTDD is
\begin{align}
    \mathcal{F}(\omega,T) = \frac{1}{\omega^2}\bigg|& 1 + (-1)^{N+1} e^{i\omega N\Delta t}e^{i\omega r}
    \\\notag&+ 2\sum^N_{j=1} (-1)^j e^{i \omega (j\Delta t+\epsilon_j)} \bigg|^2.
\end{align}
By expanding this expression and averaging over the scheduling error and remainder time distributions, we find
\begin{widetext}
\begin{align}\label{eq::av_real_time_FF}
    \langle \mathcal{F}(\omega,T) \rangle_{\text{rt}} =&\, \frac{1}{\omega^2}\bigg\{ 
    2\bigg[1 + (-1)^{N+1}\Big\langle \cos\big(\omega N\Delta t+\omega r\big)\Big\rangle_\text{rt} \bigg] \nonumber\\
    &+ 4\text{Re}\bigg[ \Big(1+(-1)^{N+1}e^{i\omega N\Delta t}\langle e^{i\omega r}\rangle_\text{rt}\Big) 
    e^{-i\omega\mu-\omega^2\sigma^2/2}\sum^N_{j=1}(-1)^j e^{-i\omega j\Delta t} \bigg] \nonumber\\
    &+ 4e^{-\omega^2\sigma^2}\sum^N_{j=1}\sum^N_{k=1}(-1)^{j+k} e^{i\omega\Delta t(j-k)} 
    +4N\big(1-e^{-\omega^2\sigma^2}\big) \bigg\},
\end{align}
where
\begin{align}
\langle \cos(\omega N\Delta t+\omega r)\rangle_\text{rt} 
    =&\, \frac{1}{\omega\Delta t}\big[\sin(\omega(N+1)\Delta t)-\sin(\omega N\Delta t) \big],\\
    \langle e^{i\omega r}\rangle_\text{rt} =&\, \frac{i}{\omega\Delta t}(1-e^{i\omega \Delta t}).
\end{align}
\end{widetext} 
The DC component of the RTDD FF, which determines the extent to which RTDD can suppress low frequency noise, is
\begin{align}\label{eq::av_real_time_FF_DC}\notag
    \langle \mathcal{F}(0,T) \rangle_{\text{rt}} =&\, 4N\sigma^2+\frac{\Delta t^2}{3}+\Delta t\mu[1-(-1)^N]\\
    &+\mu^2[1-(-1)^N]^2.
\end{align} 
Optimally suppressing low-frequency noise with RTDD will again depend on choosing a threshold time that balances the contribution from the remainder time, which scales as $\Delta t^2$, with the contribution from scheduling error, scaling as $N\sim 1/\Delta t$. 

\begin{figure}
    \centering
    \includegraphics[width=\linewidth]{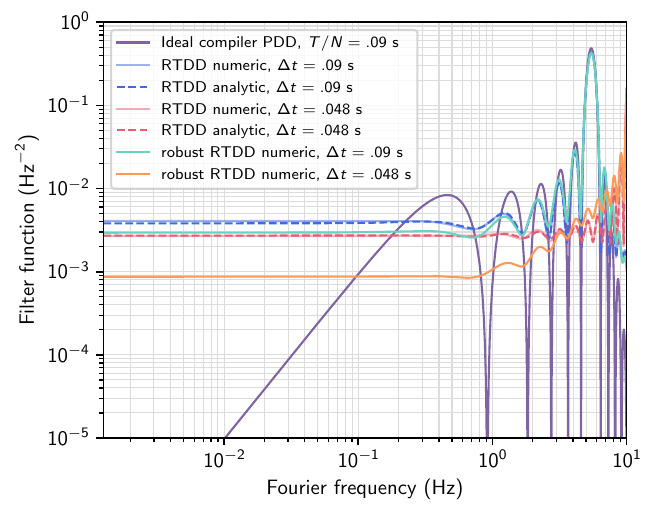}
    \caption{Filtering properties of RTDD for $\mu=2\times 10^{-4}$ s and $\sigma=4.8\times 10^{-3}$ s. FFs of RTDD with two different threshold times $\Delta t=.09$s with $N=\lfloor T/\Delta t \rfloor=11$ (blue) and $\Delta t=.048$s with $N=\lfloor T/\Delta t \rfloor=21$ (pink). Numerical FFs (solid lines) show excellent agreement with analytic FFs (dashed lines) computed via Eq. \eqref{eq::av_real_time_FF}. Unlike the FF of ideal compiler PDD with $T=1$ s and $N=11$ (purple), which decays quadratically as $\omega\rightarrow 0$, the real-time DD FFs flatten at low frequency with a DC component that increases with the threshold time. The FFs of robust RTDD for $\Delta t=.09$s (cyan) and$\Delta t=.048$s (orange) have reduced DC components due to the suppression of scheduling error. Numeric FFs are computed by averaging over 1000 realizations of scheduling error and the remainder time.}
    \label{fig:real_time_ff}
\end{figure}

\begin{figure}
    \centering
    \includegraphics[width=\linewidth]{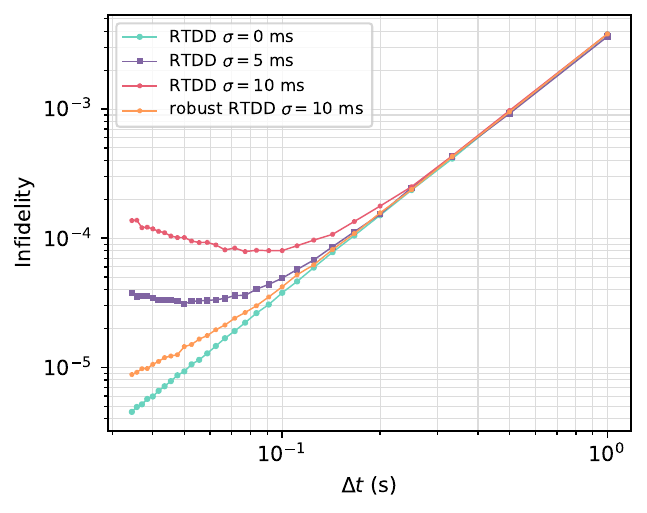}
    \caption{Infidelity versus threshold for RTDD and robust RTDD. Memory error is modeled as purely stochastic ($m=0$) with the $1/\omega^2$ PSD shown in Fig. \ref{fig:psd}. Infidelity of RTDD is plotted for scheduling error with $\mu=0$ s and varying strength set by the standard deviation: $\sigma=0$ms (cyan), $\sigma=5$ms (purple), and $\sigma=10$ms (pink). Infidelity increases monotonically with $\Delta t$ for $\sigma=0$ms, while infidelity is minimized at thresholds that balance the tradeoff between memory error and scheduling error for $\sigma=5$ms and $\sigma=10$ms. The infidelity of robust RTDD with $\sigma=10$ms (orange) approaches the infidelity of RTDD without scheduling error.}
    
    \label{fig:sq_fidelity_vs_deltaT}
\end{figure}

The low-frequency filtering properties of RTDD are illustrated in  Fig. ~\ref{fig:real_time_ff}, which plots
$\langle \mathcal{F}(\omega,T) \rangle_{\text{rt}}$ for two different threshold times, $\Delta t=.09$ s and $\Delta t=.048$ s. As expected from Eq. \eqref{eq::av_real_time_FF_DC}, the remainder error combined with scheduling error causes a finite shelf to form in the DC regime. Ideal PDD for odd $N$, in contrast, decays quadratically to zero. For the parameters we consider, the accumulation of memory error during the remainder time dominates over scheduling error. Consequently, the DC component of the RTDD FF is reduced for the smaller threshold time $\Delta t=.048$ s, despite the introduction of more scheduling error with increased $N$. Figure  \ref{fig:sq_fidelity_vs_deltaT} shows the dependence of the qubit infidelity on the threshold time for purely stochastic ($m=0$) memory error and scheduling error of varying strength. For $\sigma=5$ms and $\sigma=10$ms,  the memory error accumulated during the remainder time makes the dominant contribution to the infidelity when $\Delta t\gtrsim\,10^{-1}$ s and, accordingly, the infidelity increases with $\Delta t$. In absence of scheduling error, the infidelity monotonically increases with $\Delta t$ over the entire range. When scheduling error is present, however, infidelity is minimized at $\Delta t\approx 10^{-1}$ for $\sigma=10$ms and $\Delta t\approx 10^{-1.2}$ for $\sigma=5$ms. When $\Delta t$ decreases below these minima, the infidelity grows as scheduling error accumulates from the increased number of pulses. The optimal threshold balances the tradeoff between memory error suppression and scheduling error.

\section{Accounting for scheduling errors\label{sec:accounting}}
Using a simple protocol that can be implemented at compile time or at execution time with RTDD, we can suppress the effects of scheduling error and partially recover the performance of an ideal DD sequence.  The key to this protocol, which we term scheduling-error-robust DD,  is tracking the displacement of pulses due to scheduling error and selecting the times of subsequent pulses to cancel the contribution of scheduling error on average. Throughout this section, we consider the experimentally relevant limit where displacements due to scheduling error are much smaller than the average pulse separation, i.e., $|\mu|,\,\sigma\ll T/N$

To illustrate scheduling-error-robust DD, consider an ideal $N=2$ CPMG sequence with pulse times $t_1=T/4$ and $t_2=3T/4$. Without scheduling error, this sequence perfectly cancels coherent memory error, which we can see by evaluating the first cumulant in Eq. \eqref{eq::C1_R}
\begin{align*}
C^{(1)}(T)=&\,m\bigg[t_1-(t_2-t_1)+(T-t_2)\bigg]\\
=&\,m\bigg[\frac{T}{4}-\frac{T}{2}+\frac{T}{4}\bigg]\\
=&\,0.
\end{align*}
If the pulses occur at times $\tilde{t}_1=T/4+\epsilon_1$ and $\tilde{t}_2=3T/4+\epsilon_2$ due to scheduling error, the coherent memory error contribution is 
\begin{align*}
C^{(1)}(T)=&\,m\bigg\{\frac{T}{4}+\epsilon_1-\Big[\frac{3T}{4}+\epsilon_2-\Big(\frac{T}{4}+\epsilon_1\Big)\Big]\\&+\Big[T-\Big(\frac{3T}{4}+\epsilon_2\Big)\Big]\bigg\}\\
=&\,2\epsilon_1-2\epsilon_2.
\end{align*}
The rate of incoherent dephasing from scheduling error, which depends on the variance of this expression with respect to the scheduling error distribution, is $|K_\text{sch}^{(2)}(T)|=4m^2\sigma^2$. Suppose instead we choose the time of the second pulse to depend on the displacement of the first pulse due to scheduling error. We take the target time of the second pulse to be $u_2=3T/4+\epsilon_1$, though the pulse will be applied at the actual time $\tilde{u}_2=3T/4+\epsilon_1+\epsilon_2$ due to scheduling error. The first cumulant becomes
\begin{align*}
C^{(1)}(T)=&\,m\bigg\{\frac{T}{4}+\epsilon_1-\Big[\frac{3T}{4}+\epsilon_1+\epsilon_2-\Big(\frac{T}{4}+\epsilon_1\Big)\Big]\\&+\Big[T-\Big(\frac{3T}{4}+\epsilon_1+\epsilon_2\Big)\Big]\bigg\}\\
=&\,-2\epsilon_2.
\end{align*}
The scheduling error of the first pulse is canceled, leaving only the scheduling error of the second pulse. The rate of incoherent dephasing from scheduling error is then $|K_\text{sch}^{(2)}(T)|=2m^2\sigma^2$, reduced by half.

\subsection{Robust compile-time DD}
This strategy is easily generalized to arbitrary compile-time DD sequences with $N\geq 2$ pulses. Consider a ``base" DD sequence with pulses at target times $t_1,\ldots,t_N$ for $N\geq 2$. We can make the base DD sequence robust to scheduling errors by modifying the target pulse times as
\begin{align}
u_j=\begin{cases} t_1,& j=1,\\ 
t_j+\epsilon_{j-1}, &N\geq j>1. \end{cases}
\end{align}
Accounting for scheduling error, these pulses will actually be applied at times
\begin{align}\label{eq::u_tilde}
\tilde{u}_j=\begin{cases} t_1+\epsilon_1,& j=1,\\ t_j+\epsilon_{j-1}+\epsilon_j, &N\geq j>1. \end{cases}
\end{align}
Just like our example with $N=2$ CPMG, the timing of the subsequent pulse cancels the scheduling error contribution from the previous pulse. This is evident from the first cumulant,
\begin{align}\label{eq::C1_sch_robust}
C^{(1)}(T) =&\, m \bigg[ 2\sum^{N}_{j=1}(-1)^{j-1}\tilde{u}_j +\! (-1)^N T \bigg]\\\notag
=&\,m\int_0^Tdt R_z^z(t)+2m(-1)^{N-1}\epsilon_N,
\end{align}
where the first term on the second line vanishes for an ideal DD sequence. This expression only depends on the scheduling error that accompanies the final pulse, whereas the first cumulant for a standard, non-robust DD sequence in Eq. \eqref{eqn:cumulant_tj_epsilon} depends on the scheduling error accumulated over all $N$ pulses. 

If the memory error is purely coherent, the scheduling error cumulants and, thus, the dynamics depend exclusively on $C^{(1)}(T)$. 
Equation \eqref{eq::C1_sch_robust} shows that the scheduling error from the final pulse is the only scheduling error that contributes to $C^{(1)}(T)$. Consequently, the rate of incoherent dephasing from scheduling error is independent of the pulse number, 
\begin{align}\label{eq::K2_RT}
|K_\text{sch}^{(2)}(T)|=2m^2\sigma^2. 
\end{align}
Recall that $|K_\text{sch}^{(2)}(T)|$ increases linearly with pulse number in the $|\mu|,\sigma\ll T/N$ limit for a standard DD sequence, as seen in Eq.~\eqref{eqn:coherent_error_stats_var}. Note, however, that the coherent contribution to $K_\text{sch}^{(1)}(T)$ is
generally non-vanishing for scheduling-error-robust DD, as it is proportional to the mean of Eq. \eqref{eq::C1_sch_robust},
\begin{align}\label{eq::mean_C1_RT}
\langle C^{(1)}(T)\rangle_\text{sch}=-2m(-1)^{N-1}\mu. 
\end{align}
For a standard DD sequence subject to scheduling error, in contrast, the coherent contribution given in Eq. \eqref{eqn:coherent_error_stats_mean} vanishes for even $N$. When the memory error is coherent, the performance of scheduling-error-robust DD compared to standard DD will ultimately be set by the relative magnitudes of $\mu$ and $\sigma$.

To analyze the impact of scheduling-error-robust DD when the memory error is stochastic, we examine the incoherent contribution to $K_\text{sch}^{(1)}(T)$, which is proportional to $\langle C^{(2)}(T)\rangle_\text{sch}$.
From Eq. \eqref{eq::C2_FF_av}, recall that  $\langle C^{(2)}(T)\rangle_\text{sch}$ depends on an overlap integral between the PSD of $\beta(t)$ and the FF averaged over the distribution of scheduling error. Using an approach similar to Sec. \ref{ssec:stoch_env}, we show in Appendix \ref{sec::robust_FF} that the average FF of a scheduling-error-robust DD sequence is 
\begin{widetext}
\begin{align}\label{eq::robust_FF}
\big\langle F(\omega,T)\big\rangle_\text{sch}
=&\,\frac{1}{\omega^2}\bigg\{\Big|1+(-1)^{N+1}e^{i\omega T}\Big|^2
+4N\notag
+4\text{Re}\Big[\big(1+(-1)^{N+1}e^{i\omega T}\big)\sum_{j=1}^N(-1)^je^{-i[t_j+(2-\delta_{j,1})\mu]\omega-(1-\frac{\delta_{j,1}}{2})\sigma^2\omega^2}\Big]\\
&-\! 8\text{Re}\Big[\sum_{j=1}^{N-1}e^{i(t_{j+1}-t_{j}+\delta_{j,1}\mu)\omega-(1-\frac{\delta_{j,1}}{2})\sigma^2\omega^2}\Big]\!+\! 8\text{Re}\Big[\sum_{j'=1}^{N-2}\sum_{j=j'+2}^N(-1)^{j+j'}e^{i(t_{j}-t_{j'}+\delta_{j',1}\mu)\omega-(2-\frac{\delta_{j',1}}{2})\sigma^2\omega^2}\Big]\bigg\} .
\end{align}
\end{widetext}
Similar to the scheduling error-averaged FF in Eq. \eqref{eq::FF_av_DC}, the DC component can be expressed in terms of the ideal FF without scheduling error plus an additive bias,
\begin{align}\label{eq::FF_av_DC_robust}
\langle \mathcal{F}(0,T) \rangle_{\text{sch}} 
    =&\, \mathcal{F}_\text{ideal}(0,T)+ 4[\sigma^2+\mu^2]
    \\\notag&\,+4\mu(-1)^{N-1}\int_0^Tdt\, R_{z,\,\text{ideal}}^z(t).
\end{align}
Unlike Eq. \eqref{eq::FF_av_DC}, however, the scheduling error-dependent terms do not scale linearly with $N$, indicating that scheduling error is suppressed.

The effectiveness of scheduling-error-robust DD in suppressing low-frequency noise is illustrated in Fig. \ref{fig:timing_error_ff}, which shows the filter functions of $N=6$ CPMG with and without scheduling error. The DC component of the scheduling-error-robust FF is smaller than the FF of a standard CPMG sequence by nearly an order of magnitude. While scheduling-error-robust DD is not as effective in suppressing low frequency noise as an ideal DD sequence without scheduling error, the reduction in the size of the DC component marks a significant improvement over standard DD. 

\begin{figure*}[t!]
    \centering
    \includegraphics[width=\linewidth]{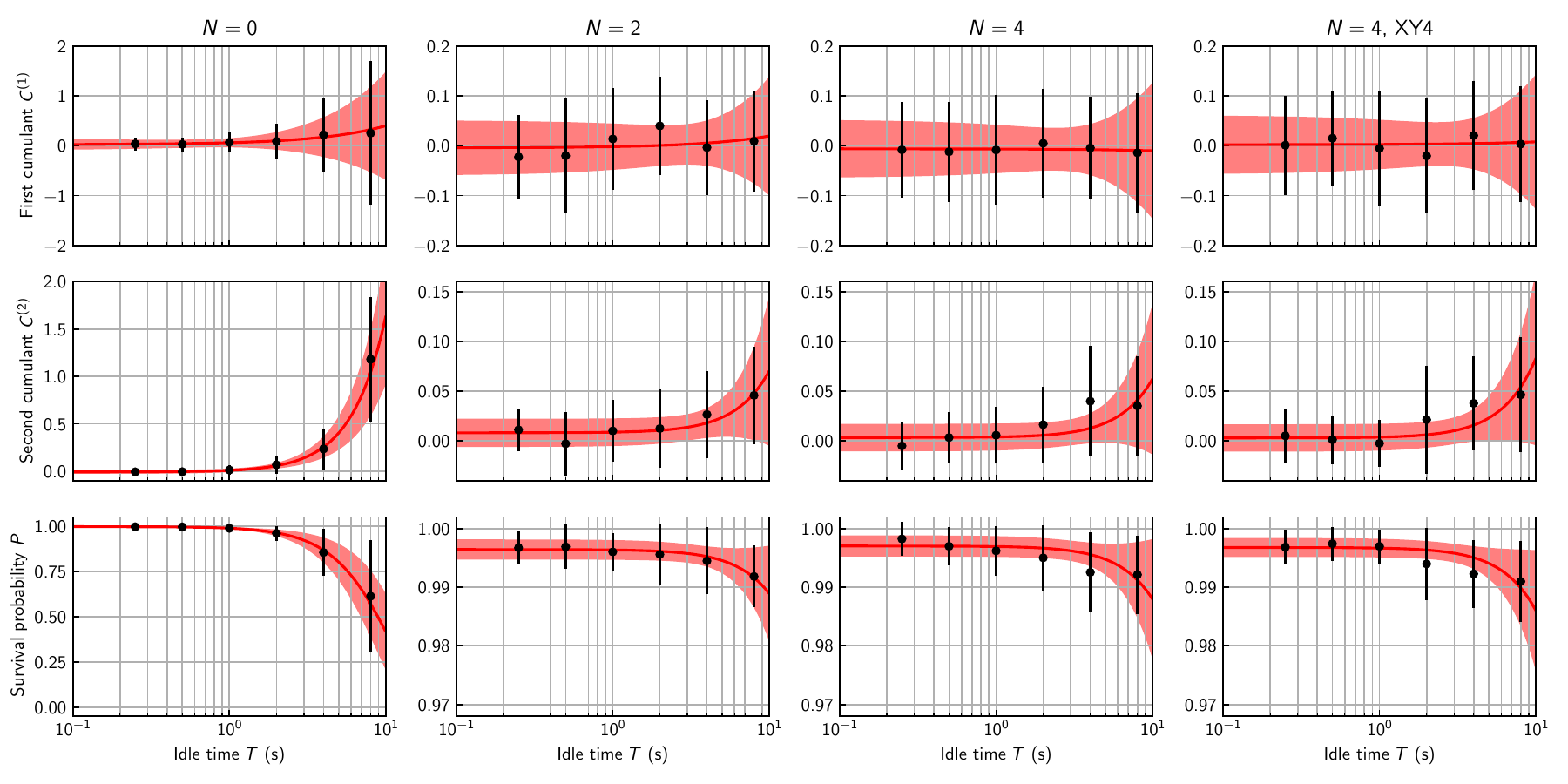}
    \caption{
        Ramsey delay experiments with DD performed on Quantinuum H2-1.
        The single-qubit survival probabilities are marginalized and averaged over 56 qubits, with error bars representing $1~\sigma$ standard deviations.
        The rightmost column corresponds to an ``XY4'' sequence which is identical to the $N=4$ CPMG sequence, except that the $j=2,4$ pulses are performed about the $\sigma^y$ axis.
    }
    \label{fig:sq_fidelity}
\end{figure*}

\subsection{Robust real-time DD}
Scheduling-error-robust DD can also be adapted to the execution-time setting of RTDD. Recall that RTDD attempts to opportunistically apply pulses at multiples of a threshold time $\Delta t$, e.g., $t_j=j\Delta t$. By tracking the scheduling error, we can make RTDD robust by instead applying pulses opportunistically at the target times
\begin{align}
u_j=\begin{cases} \Delta t,&j=1,\\
j\Delta t+\epsilon_{j-1},& j>1.\end{cases}
\end{align}
Accounting for scheduling error, the pulses will instead be applied at times
\begin{align}
\tilde{u}_j=\begin{cases} \Delta t+\epsilon_1,&j=1,\\
j\Delta t+\epsilon_{j}+\epsilon_{j-1},& j>1.\end{cases}
\end{align}
Like the compile-time case, the timing of the pulses is designed to cancel the scheduling error that has entered through the previous pulses, though the RTDD adds an additional memory error contribution from the remainder time.

We again assess the performance of this strategy by examining the cumulants and FF. The first cumulant shows that all scheduling error is canceled, except for the scheduling error that accompanies the final pulse,
\begin{align}
C^{(1)}(T)=\frac{\Delta t}{2}\big[1-(-1)^N\big]+(-1)^Nr+2(-1)^{N-1}\epsilon_N.
\end{align}
The coherent contribution to $K_\text{rt}^{(1)}(T)$ and the second real-time cumulant, which follow from $C^{(1)}(T)$, are
\begin{align}
\langle C^{(1)}(T)\rangle_\text{rt}=\frac{m\Delta t }{2}+2m(-1)^{N-1}\mu,
\end{align}
\begin{align}
|K_\text{rt}^{(2)}(T)|=\frac{m^2\Delta t^2}{24}+2m^2\sigma^2.
\end{align}
As expected, the dephasing rate set by $|K_\text{rt}^{(2)}(T)|$ is independent of $N$, unlike standard RTDD. The average DC component of the FF, which follows from Eq. \eqref{eq::FF} with $\omega=0$, is
\begin{align}
\langle \mathcal{F}(0,T)\rangle_\text{rt}=4[\sigma^2+\mu^2]+\frac{\Delta t^2}{3}-2\Delta t\mu(-1)^N.
\end{align}
Unlike standard RTDD, the DC component does not scale linearly with $N$.

The reduced impact of scheduling errors on robust RTDD is illustrated in Figs. \ref{fig:real_time_ff} and \ref{fig:sq_fidelity_vs_deltaT}. Compared to standard RTDD, Fig. \ref{fig:real_time_ff} shows that robust RTDD produces an FF with a reduced DC component. The reduction in the DC component is particularly noticeable for the smaller threshold time $\Delta t=.048$ s, as a smaller threshold time implies more pulses. 
In Fig. \ref{fig:sq_fidelity_vs_deltaT}, which plots qubit infidelity versus RTDD threshold, the infidelity of robust RTDD decreases with the threshold $\Delta t$. For standard RTDD with nonzero scheduling error, in contrast, infidelity decreases with $\Delta t$ until it reaches a minimum, after which it increases due to the accumulation of scheduling error. Since robust RTDD suppresses scheduling error from all pulses except the final one, error from the remainder time can be reduced by making the threshold smaller without the accumulation of scheduling error from higher pulse numbers.

\section{Experimental Results}\label{sec::experimental_results}

Next, we present experimental demonstrations of compile-time DD techniques.
We perform single-qubit Ramsey experiments simultaneously on all 56 qubits of the Quantinuum H2-1 quantum computer and average the survival probabilities to obtain single-qubit fidelities for various CPMG sequences. To apply DD pulses, the 56 qubits must be transported in and out of the 8 gate zones. As a result, scheduling error will affect the the timing of the pulses. For each DD sequence, the qubits are prepared in the $+1$ eigenstate of $\sigma^x$,  evolved for a time $T$, and then measured in the $\sigma^\phi = \cos(\phi) \sigma^x+\sin(\phi)\sigma^y$ basis where $\phi$ is a programmable ``readout'' phase. 

The expectation value of $\sigma^\phi$ can be expressed in terms of the scheduling error cumulants in Eqs. (\ref{eq::K1}) and (\ref{eq::K2}). Here, we use the shorthand $C^{(1)}(T)\equiv\langle C^{(1)}(T)\rangle_\text{sch}$ to denote the coherent component of $K_\text{sch}^{(1)}(T)$ and $C^{(2)}(T)\equiv\langle C^{(2)}(T)\rangle_\text{sch}+\text{var}[C^{(1)}(T)]$ to denote the components of $K_\text{sch}^{(1)}(T)$ and $K_\text{sch}^{(2)}(T)$ from incoherent dephasing. The expectation value is then
\begin{align}
    \langle\sigma^\phi\rangle(T) = \cos\big[C^{(1)}(T)-\phi\big]e^{-C^{(2)}(T)/2}~.
\end{align}
Performing measurements over four readout phases $\phi\in\{-\pi, \pi/2, 0, \pi/2\}$ at each time $T$ then allows for unambiguous reconstruction of the cumulants in a manner that is robust to small pulse errors and leakage out of the qubit subspace
\begin{align}
\hat{C}^{(1)}(T)&=\text{arg} \sum_\phi e^{i \phi} \langle\hat{\sigma}^\phi\rangle(T),\\
\hat{C}^{(2)}(T)&=-2 \log \Big| \frac{1}{2}\sum_\phi e^{i \phi} \langle\hat{\sigma}^\phi\rangle(T)\Big|~.
\end{align}
We then quantify the effectiveness of a given DD sequence by reconstructing the survival probability according to
\begin{align}
    P = \frac{1}{2} + \frac{1}{2} \cos\big[\hat{C}^{(1)}(T)\big]e^{-\hat{C}^{(2)}(T)/2}.
\end{align}

In Fig.~\ref{fig:sq_fidelity}, we show the inferred cumulants and survival probabilities for $N=0,2,4$ CPMG sequences at idle times $T$ ranging from 250 ms to 8 s. The experimental data show significant improvements in survival probabilities for $N>0$ as compared to $N=0$, indicating that the noise is concentrated at low frequency.  For the maximum pulse number of $N=4$ considered here, the survival probability does not decrease by a statistically significant amount with increasing $N$ due to the accumulation of scheduling error. No significant difference is observed between the $N=2$ and $N=4$ sequences, however, indicating that fidelity is not improved by increasing the pulse number.  Additionally, no significant difference is observed between 
$N=4$ CPMG with pulses applied about $\sigma^x$ and $N=4, XY4$, in which the pulse axis alternates between $\sigma^x$ and $\sigma^y$. This demonstrates that the noise is dominated by longitudinal magnetic field fluctuations, in accordance with our model.

\section{Conclusions\label{sec:conclusion}}
In this work, we presented analytic, numeric, and experimental results to develop a DD strategy for trapped-ion QCCDs.
We reviewed two leading sources of errors on trapped-ion QCCDs, memory errors and scheduling-induced timing errors.
We developed an analytical model to quantify the impact of scheduling errors on the performance of DD. Our model demonstrates that scheduling error limits the ability of DD to suppress low frequency dephasing noise, producing dephasing rates that scale linearly with pulse number.

In addition to considering DD injected at compile-time, we introduced RTDD, a protocol for applying DD at execution time. For quantum computers that schedule pulses at runtime, such as Quantinuum Helios~\cite{ransford2025helios98qubittrappedionquantum}, RTDD opportunistically applies a DD pulse after an ion has been idle for a specified threshold time, $\Delta t$. Using our analytic framework, we found that the fidelity achievable with RTDD is ultimately set by the tradeoff between memory error that accumulates for larger $\Delta t$ and scheduling error that accumulates for smaller $\Delta t$. For both compile-time and RTDD, we developed scheduling-error-robust DD where the placement of future DD pulses is updated based on the scheduling errors in earlier DD pulses. This procedure suppresses the accumulation of scheduling error, producing dephasing rates that are independent of pulse number.

\section{Acknowledgments}
We thank Yi-Hsiang Chen, Jason Dominy, Matthew DeCross, John Gaebler, and Aaron Hankin for useful discussions and feedback.

\appendix

\begin{widetext}

\section{Distributions of Pulse Times}\label{sec::pulse_time_distributions}
Figures \ref{fig:pulse_timing_linear} and \ref{fig:pulse_timing_PT} show numerical simulations of the distributions of stochastic pulse timings in the limits $T/N\gg \sigma$ and $T/N\ll \sigma$, respectively. In the $T/N\ll \sigma$ limit with $N\gg1$, shown in Fig. \ref{fig:pulse_timing_PT}, the distribution of pulse times approaches a uniform distribution.

\begin{figure}[H]
    \centering
    \includegraphics[width=0.8\linewidth]{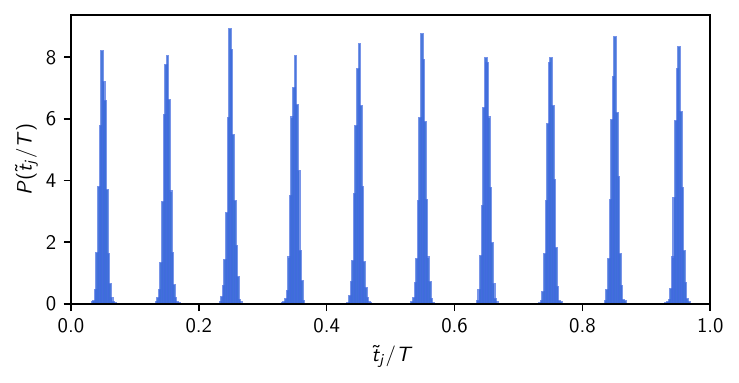}
    \caption{Distribution of pulse times subject to scheduling error for CPMG with $N=10$, $T=1$ s, $\mu=2\times 10^{-4}$ s, and $\sigma=4.8\times 10^{-3}$ s. For these parameters, $T/N\gg |\mu|,\,\sigma$ and the pulses remain localized near their ideal values, $t_j/T=\frac{2j-1}{2N}$, with high probability.
    }
    \label{fig:pulse_timing_linear}
\end{figure}

\begin{figure}[H]
    \centering
    \includegraphics[width=0.8\linewidth]{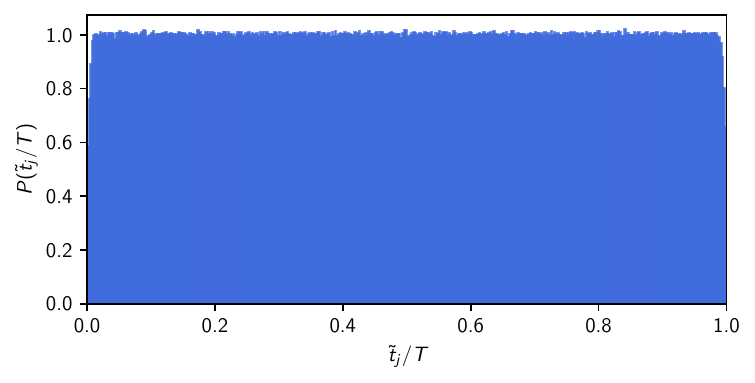}
    \caption{Distribution of pulse times subject to scheduling error for CPMG with $N=5000$, $T=1$ s, $\mu=2\times 10^{-4}$ s, and $\sigma=4.8\times 10^{-3}$ s. For these parameters, $T/N\ll \sigma$ and the distribution of pulse times approaches a uniform distribution.
    }
    \label{fig:pulse_timing_PT}
\end{figure}

\section{Average FF for scheduling-error-robust DD}\label{sec::robust_FF}
Consider a scheduling-error-robust DD sequence applied over a single realization of scheduling error with pulses occurring at $\tilde{u}_1,\ldots,\,\tilde{u}_N$, as given in Eq. \eqref{eq::u_tilde}. The FF of this sequence follows from Eq. \eqref{eqn:dd_ff},
\begin{align}
F(\omega,T)=&\,\frac{1}{\omega^2}\bigg| 1+(-1)^{N+1}e^{i\omega T}+2\sum_{j=1}^N(-1)^je^{i\tilde{u}_j\omega} \bigg|^2\notag\\
=&\,\frac{1}{\omega^2}\bigg\{\Big|1+(-1)^{N+1}e^{i\omega T}\Big|^2+4\text{Re}\Big[\big(1+(-1)^{N+1}e^{i\omega T}\big)\sum_{j=1}^N(-1)^je^{-i\tilde{u}_j\omega}\Big]
+4N+ 8\text{Re}\Big[\sum_{j>j'=1}^N(-1)^{j+j'}e^{i(\tilde{u}_j-\tilde{u}_{j'})\omega}\Big]\bigg\}. 
\end{align}
By taking the average of the FF over the distribution of scheduling errors, we find
\begin{align}
\expect{F(\omega,T)}_\text{sch}
=&\,\frac{1}{\omega^2}\bigg\{\Big|1+(-1)^{N+1}e^{i\omega T}\Big|^2+4\text{Re}\Big[\big(1+(-1)^{N+1}e^{i\omega T}\big)\sum_{j=1}^N(-1)^j\expect{e^{-i\tilde{t}_j\omega}}_\text{sch}\Big]
+4N\\&\,+ 8\text{Re}\Big[\sum_{j>j'=1}^N(-1)^{j+j'}\expect{e^{i(\tilde{t}_j-\tilde{t}_{j'})\omega}}_\text{sch}\Big]\bigg\} \label{eq::avFF_app}.
\end{align}
We can again use a cumulant expansion to evaluate the characteristic functions,
\begin{align}
\expect{e^{-i\tilde{t}_j\omega}}_\text{sch}=&\,\begin{cases} e^{-i(t_1+\mu)\omega-\frac{1}{2}\sigma^2\omega^2},&j=1,\\
e^{-i(t_j+2\mu)\omega-\sigma^2\omega^2},&N\geq j>1,
\end{cases}\\ \notag\\
=&\, e^{-i[t_j+(2-\delta_{j,1})\mu]\omega-(1-\delta_{j,1}/2)\sigma^2\omega^2}
\end{align}
and
\begin{align}
\expect{e^{i(\tilde{t}_j-\tilde{t}_{j'})\omega}}_\text{sch}=&\,\begin{cases} e^{i(t_2-t_1+\mu)\omega-\frac{1}{2}\sigma^2\omega^2},& j'=1,\,j=2,\\
e^{i(t_{j'+1}-t_{j'})\omega-\sigma^2\omega^2},&N>j'>1,\, j=j'+1,\\
e^{i(t_j-t_1+\mu)\omega-\frac{3}{2}\sigma^2\omega^2}, &j'=1, N\geq j>2,\\
e^{i(t_j-t_{j'})\omega-2\sigma^2\omega^2}, &N-1>j'>1,\,j>j'+1,
\end{cases}\\
=&\,\begin{cases} e^{i(t_{j'+1}-t_{j'}+\delta_{j',1}\mu)\omega-(1-\delta_{j',1}/2)\sigma^2\omega^2},&N>j'\geq 1,\, j=j'+1,\\
e^{i(t_{j}-t_{j'}+\delta_{j',1}\mu)\omega-(2-\delta_{j',1}/2)\sigma^2\omega^2}, &N-1>j'\geq 1,\, j>j'+1.
\end{cases}
\end{align}
By substituting these expressions into Eq. (\ref{eq::avFF_app}), we obtain an explicit expression for the average FF
\begin{align}
\expect{F(\omega,T)}_\text{timing}
=&\,\frac{1}{\omega^2}\bigg\{\Big|1+(-1)^{N+1}e^{i\omega T}\Big|^2+4\text{Re}\Big[\big(1+(-1)^{N+1}e^{i\omega T}\big)\sum_{j=1}^N(-1)^je^{-i[t_j+(2-\delta_{j,1})\mu]\omega-(1-\delta_{j,1}/2)\sigma^2\omega^2}\Big]
\notag\\&\,+4N- 8\text{Re}\Big[\sum_{j=1}^{N-1}e^{i(t_{j+1}-t_{j}+\delta_{j,1}\mu)\omega-(1-\delta_{j,1}/2)\sigma^2\omega^2}\Big]\notag\\
&\,+ 8\text{Re}\Big[\sum_{j'=1}^{N-2}\sum_{j=j'+2}^N(-1)^{j+j'}e^{i(t_{j}-t_{j'}+\delta_{j',1}\mu)\omega-(2-\delta_{j',1}/2)\sigma^2\omega^2}\Big]\bigg\}.
\end{align}

\end{widetext}

\bibliography{refs}

\end{document}